\documentclass{xarch}
\title{\vspace*{-5cm}
         \hfill {\normalsize\rm Preprint HEN-458}\\
         \vspace*{4cm}
         Tests of the Electroweak Sector of the Standard Model}
\ShortTitle{Tests of the EW SM}
\author{\speaker{Sijbrand de Jong}%
              \thanks{
              \ Plenary presentation at the
              HEP2005 International Europhysics Conference on High Energy Physics,
              EPS (July 21st-27th, 2005) in Lisboa, Portugal.
              }\\
              Institute for Mathematics, Astrophysics and Particle Physics\\
              Radboud University Nijmegen and NIKHEF\\
              Toernooiveld 1\\
              6525 ED Nijmegen\\
              The Netherlands\\
              Email: s.dejong@science.ru.nl}
\FullConference{\ }

\abstract{The Electroweak sector of the Standard Model is reviewed and best fits
are presented for its free parameters based on currently available experimental
tests.
The Standard Model remains an excellent descriptions of the
available experimental data.
The preferred mass range of the still elusive Higgs boson in the
Standard Model is $114<m_{\mathrm{H}}<219$~GeV
at the 95\% Confidence Level.
A Standard Model Higgs in this mass range is likely to be observed
in the years 2007--2010, either at the Tevatron or at the LHC.
}
\begin{document}

\section{Introduction}
\vspace*{-3mm}
The Standard Model (SM) of electroweak~\cite{GWSmodel}
and strong interactions~\cite{QCDmodel} is an extremely
successful theory that describes the relevant experimental data
in detail. In this review we will focus on the electroweak sector of
the SM. The strong interaction sector will be discussed in contributions
by Greenshaw~\cite{Greenshaw} and Davies~\cite{QCDonthelattice}.
Flavour physics in the quark sector is treated by
Branco~\cite{Branco} and Shune~\cite{Shune},
while neutrino physics and lepton flavour mixing is covered by
Klein~\cite{Klein} and Sanchez~\cite{Sanchez} in these proceedings.

Despite its elegant principles as a gauge theory the SM is not a trivial structure.
Try to write
down the Lagrangian after Symmetry breaking~\cite{VeltmanSM} !
And this is only part
of the story, especially higher orders in perturbation theory make 
everything connected to everything else.

The hyper charge and weak isospin part of the EW symmetry come
with separate couplings strengths,\footnote{
Rather than using the conventional term {\em coupling constant}
I will use {\em coupling strength} or just {\em coupling} in
recognition of the fact that these parameters are not constant, as
will be shown in the remainder of this proceeding.
} $g^{\prime}$ and $g$.
Instead of the coupling strengths $g^{\prime}$ and $g$ other pairs of
parameters can also be used, such as the four fermion
coupling $G_F=g^2/4\sqrt{2}M_{\mathrm{W}}^2$
and the weak mixing angle $\theta_w=\tan^{-1}(g^{\prime}/g)$\footnote{
Since the coupling strengths in theory dependent both on the order
they are calculated at and at the renormalisation scheme used,
especially for $\sin^2\theta_w$ various definitions are around.
I will conform myself to the PDG notation~\cite{PDGsin2theta}
of these variants.}, or other independent pairs.

There are three ElectroWeak (EW) gauge bosons coupling to fermions:
photon to fermions which is purely vector and has strength
$g\sin\theta_w$,
W~boson to fermions which is purely vector minus axial-vector with
strength $g/2\sqrt{2}$, and
Z~boson to fermions which is a well defined mix of vector and axial
vector couplings with strength $g/2\sin\theta_w$.
When ignoring the coupling strength, the structure in terms of vector and
axial-vector components of the vertices can be written as
$\left( v_{Vf}-a_{Vf}\gamma^5\right)\gamma^{\mu}$,
where I use the symbols
$v_{Vf}$ and $a_{Vf}$ for vector and axial-vector
coupling coefficients of the vector boson $V$ to fermion species $f$.

In the SM the vector and axial-vector couplings for vector-boson
fermion interactions can all be expressed in the charges of the fermions:\\[-4mm]
\[
\begin{array}{lll}
v_{\gamma f} = Q_f \hspace*{1cm}&
v_{\mathrm{W}f}= T_{3f} \hspace*{2cm}&
v_{\mathrm{Z}f}= T_{3f}-2Q_f \sin^2\theta_w\\
a_{\gamma f} = 0    &
a_{\mathrm{W}f}=-T_{3f} &
a_{\mathrm{Z}f}= -T_{3f}\\
\end{array}
\vspace*{-1.5mm}
\]
These are the couplings in the Lagrangian and would be the measured couplings
if only lowest order effects in the couplings are taken into account.
The effective couplings, i.e.\ those that are measured,
include effects of higher orders and become dependent on each other and on
all other parameters in the theory, such as masses and charges.
Since the relative strength of higher order contributions depend on the distance
or energy scale, all these parameters as they are measured will depend on
the energy scale at which they are measured.

Where at first sight this may seem nothing but trouble, this notion can also be
turned around. Measurements of the SM couplings can be used to predict,
e.g.\ masses of particles, as was successfully done in case of the top quark.
Presently such attention goes to the SM Higgs boson mass.

\vspace*{-3mm}
\section{Electromagnetic Interactions}
\vspace*{-3mm}
The coupling between
the photon and fermions is the realm of Quantum ElectroDynamics (QED),
which is part of the SM and which is known as the most precise theory around.
The QED couplings constant is defined as
$\alpha_{\mathrm{QED}}=e^2/4\pi=g^2\sin^2\theta_w/4\pi$.
Currently one of the major players in the field of SM precision measurements
is LEP, with the ALEPH, DELPHI, L3 and OPAL experiments,
that collected each about 800~pb$^{-1}$ of e$^+$e$^-$ data at
energies between $m_{\mathrm{Z}}$ and 209 GeV. Also important is the
SLD experiment at the SLC with a sample of 350000 Z bosons
with polarized beams.
Data collection of these experiments has been stopped, the analyses are
still ongoing and nearly finished.
As we will see, significant inputs to SM tests in the QED sector are also provided 
by other experiments.

\begin{figure}[btp]
\begin{minipage}[t]{6.5cm}
\centerline{\includegraphics[height=5cm]{pictures/l3_compton.eps}}
\caption[Compton scattering cross section]{
          \label{l3compton}
          Compton Scattering cross section as measured by the 
          L3 collaboration~\cite{l3compton}.
           }
\end{minipage}
\hfill
\begin{minipage}[t]{8cm}
\centerline{\includegraphics[height=5cm]{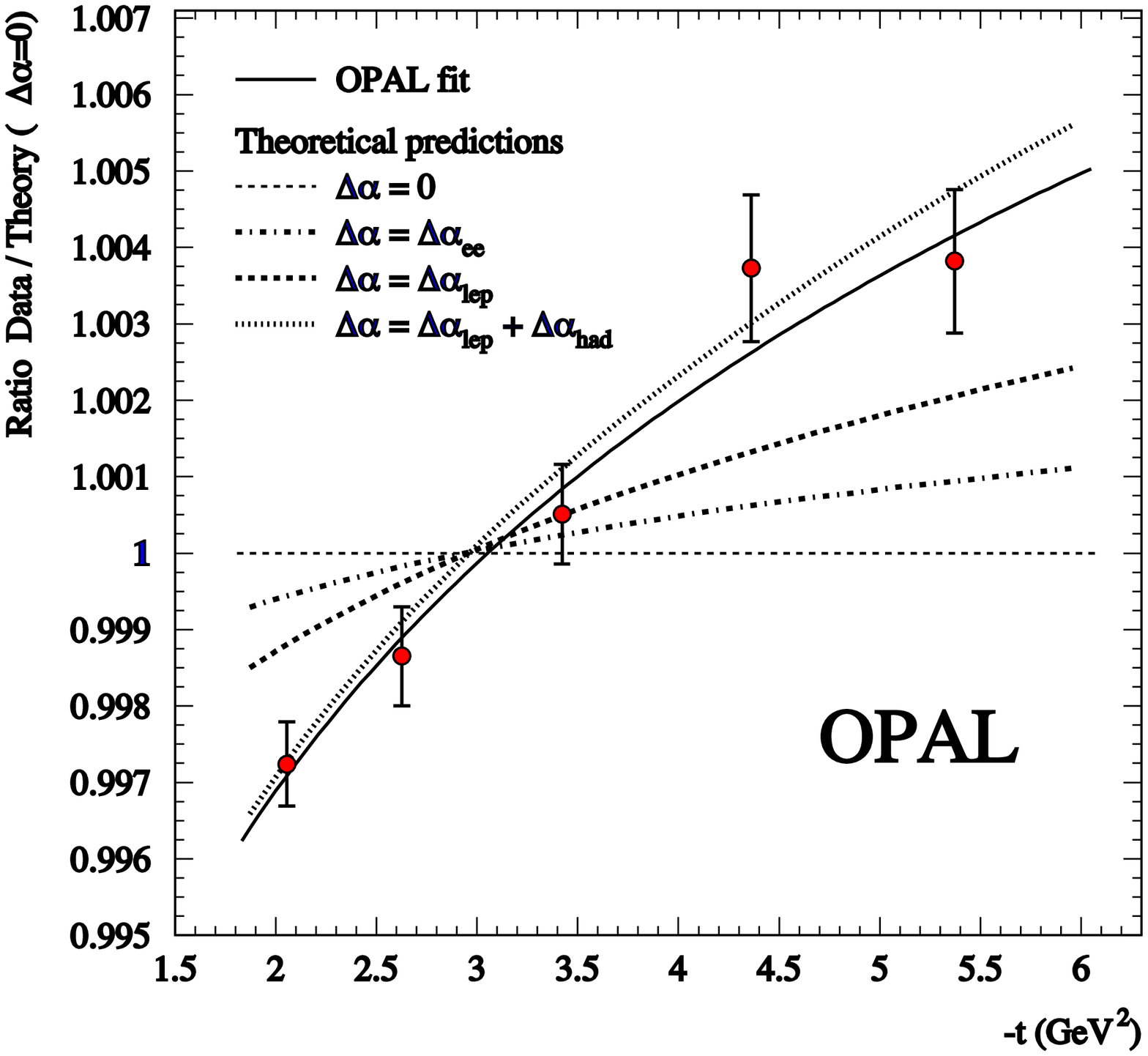}}
\caption[Running of QED coupling constant]{
          \label{opalrunalpha}
          Running of $\alpha_{QED}$ demonstrated by the OPAL
          collaboration~\cite{opalrunalpha}. Plotted is the ratio of the
          measured interaction strength divided by the expectation without
          running of the coupling constant.
          }
\end{minipage}
\vspace*{-4mm}
\end{figure}
As a demonstration of the vector character of QED, the Compton scattering
(e$^{\pm}+\gamma$$\rightarrow$e$^{\pm}+\gamma$)
cross section at high energy as measured by L3 and perfectly fitted by the SM
is displayed in Fig.~\ref{l3compton}~\cite{l3compton}.

\begin{figure}[t]
\begin{minipage}[b]{7.5cm}
\centerline{\includegraphics[height=6cm]{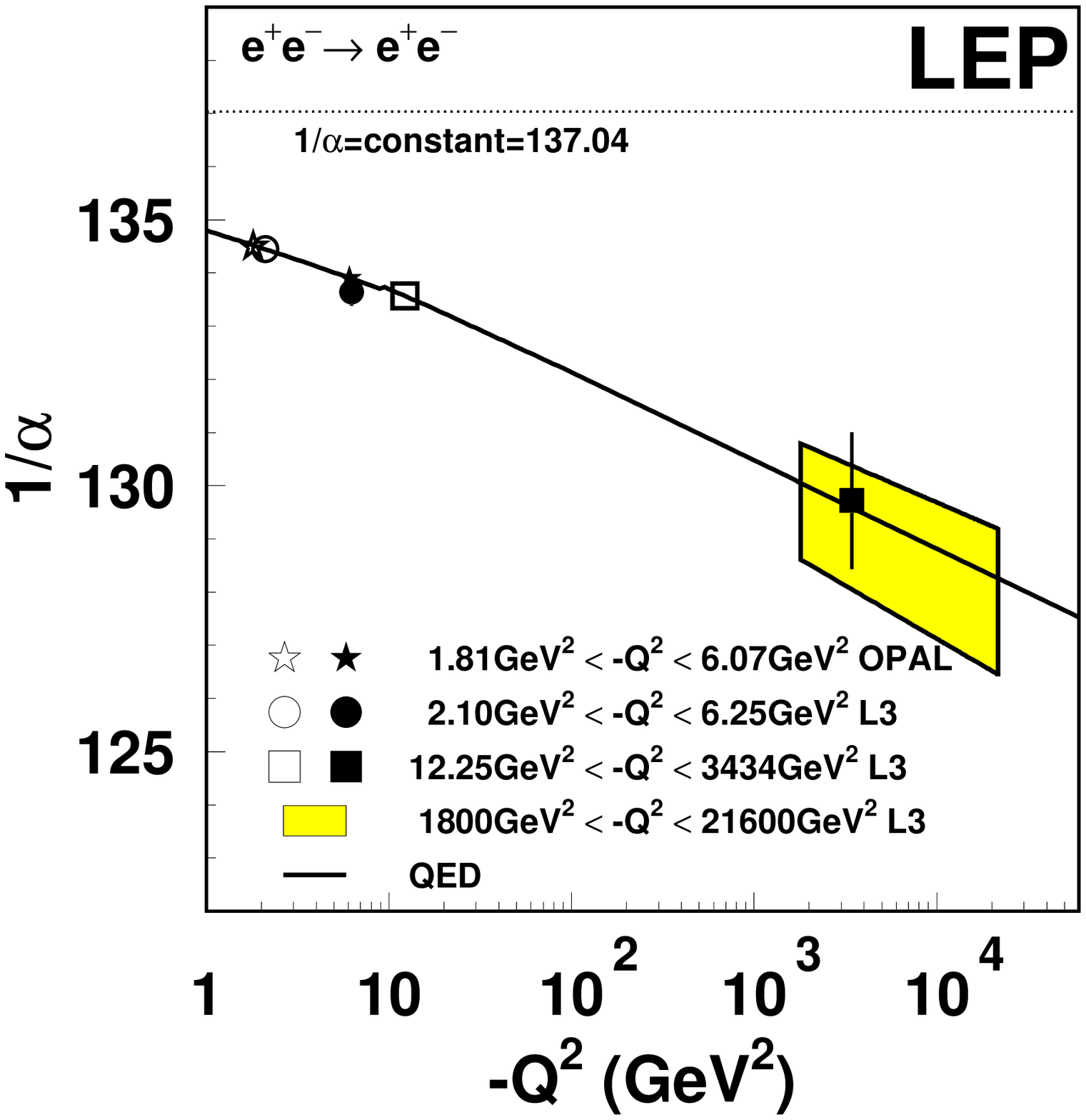}}
\caption[Running of alpha QED]{
          \label{srunalpha}
          The value of the reciprocal QED coupling constant over a large range of $-Q^2$
          values. See~\cite{srunalpha} for further details on this figure.
           }
\end{minipage}
\hfill
\begin{minipage}[b]{7cm}
\begin{tabular}{c|r|l}
$(a_{\mu}^{\mathrm{exp}}-a_{\mu}^{\mathrm{SM}})\times 10^{11}$ & $\sigma$ &
Model Ref. \\
\hline
235 (91) & 2.6 & \cite{Hoecker04} (e$^+$e$^-$) \\
221 (108) & 2.1 & \cite{Jegerlehner03} (e$^+$e$^-$) \\
235 (113) & 2.1 & \cite{Ezhela05} (e$^+$e$^-$) \\
245 (91) & 2.7 & \cite{Hagiware03} (e$^+$e$^-$) \\
225 (81) & 2.8 & \cite{deTroconiz04} (e$^+$e$^-$) \\
62 (87) & 0.7 & \cite{Davier03} ($\tau$) \\
142 (81) & 1.8 & \cite{deTroconiz04} (e$^+$e$^-$, $\tau$)
\end{tabular}
\vspace*{7mm}
\caption[Muon anomalous moment]{
               \label{amu}
               Difference of the muon anomalous magnetic moment
               predicted the theoretical models indicated in the
               third column and the measurement~\cite{BNLE821}.
               The second column gives the number of
               standard deviations~\cite{Passera}. 
              }
\end{minipage}
\vspace*{-4mm}
\end{figure}

In Fig.~\ref{opalrunalpha} the running of $\alpha_{QED}$
is demonstrated at Q$^2$ values between 2 and 6 GeV$^2$ in the
regime of space-like momentum transfer in small angle Bhabha scattering
(e$^{+}+$e$^{-}$$\rightarrow$e$^{+}+$e$^{-}$)
by the OPAL collaboration~\cite{opalrunalpha}.
This measurement confirms an earlier L3 measurement~\cite{l3runalpha}
with more precision and detail.
Although tricky (it is easy to have this measurement make a reference to itself),
the running of $\alpha_{QED}$ can be demonstrated over
a larger Q$^2$ range in the $s$-channel too.
In Fig.~\ref{srunalpha} various measurements from OPAL~\cite{opalrunalpha}
and L3~\cite{l3runalpha,l3runalphanew} are
displayed that are reinterpreted using the relation
$\alpha_{\mathrm{QED}}(Q^2)=
  \alpha_{\mathrm{QED}}(0)/
  (1-C\,\Delta\alpha_{\mathrm{QED}}(Q^2))$
as was done in a contribution to this conference by S.~Mele~\cite{srunalpha}.

\begin{figure}[bp]
\vspace*{-3mm}
\begin{minipage}[t]{5.5cm}
\centerline{\includegraphics[height=6.5cm]{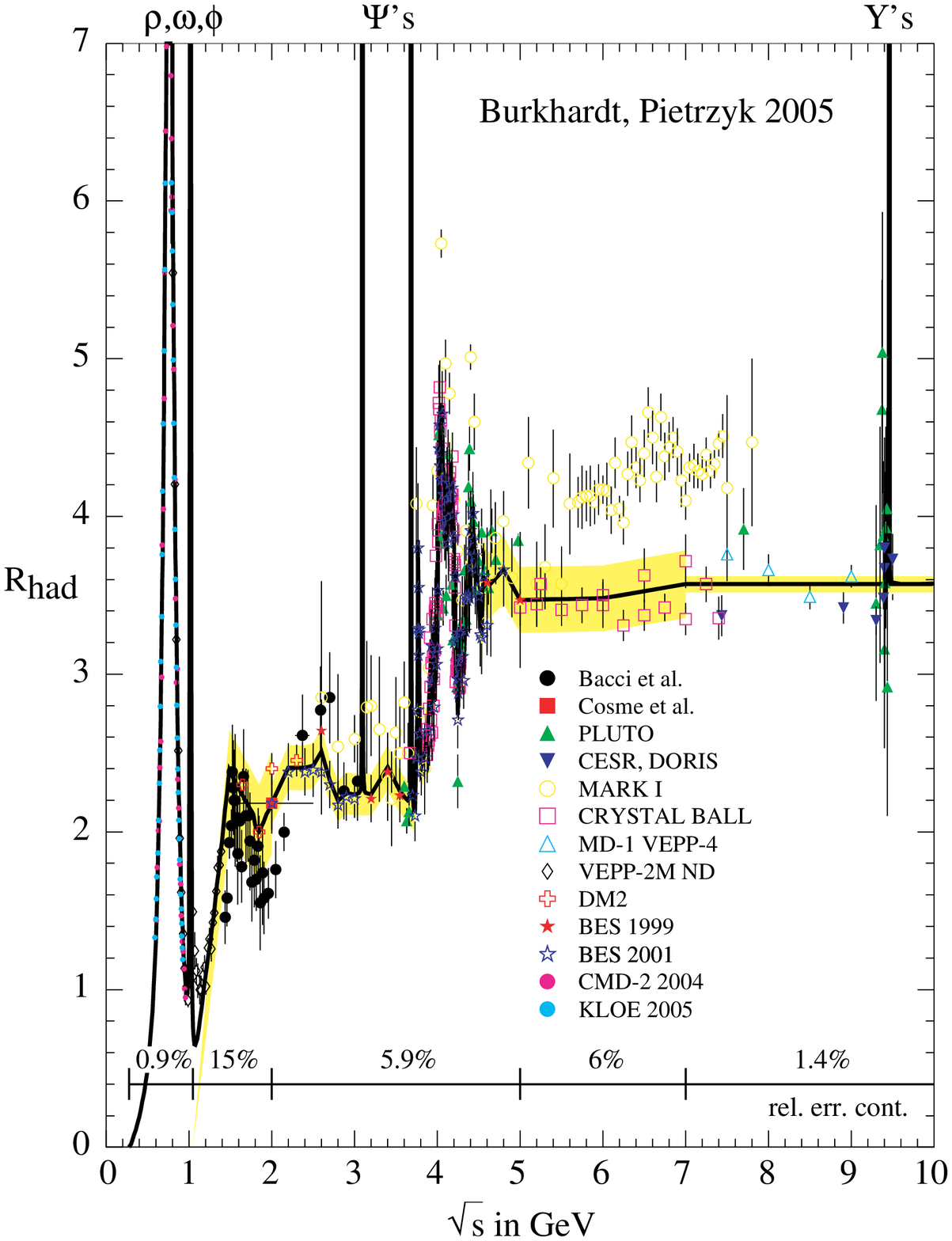}}
\caption[Alpha QED hadronic correction]{\label{alphaqed}
          Ratio of hadronic to mu-pair cross section in e$^+$e$^-$ annihilation
          as used in~\cite{burkhardt} to derive the hadronic correction to the
          running of $\alpha_{QED}$.
         }
\end{minipage}
\hfill
\begin{minipage}[t]{9cm}
\centerline{\includegraphics[height=6.5cm]{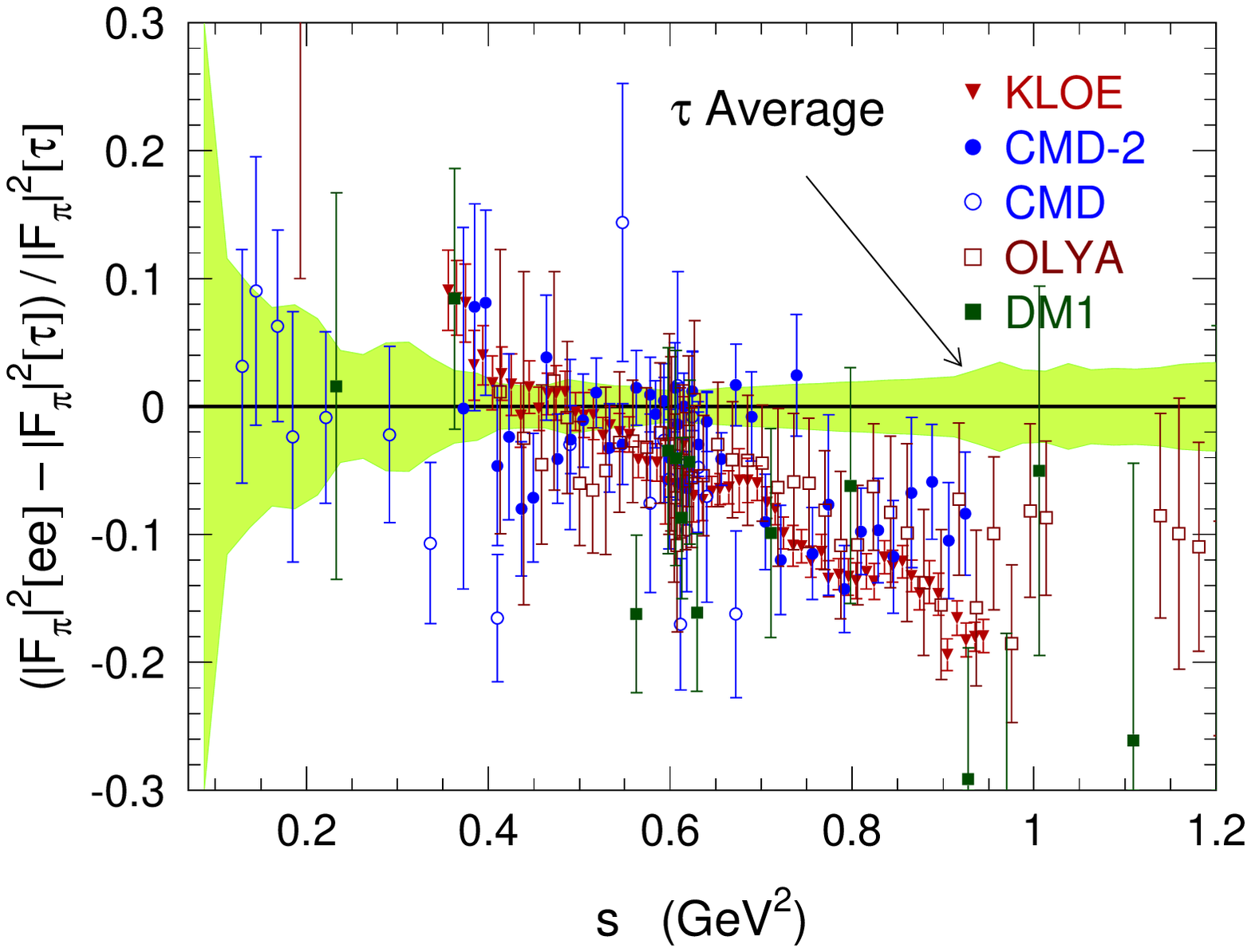}}
\caption[ee and $\tau$ spectral function comparison]{
          \label{specfun}
          e$^+$e$^-$ spectral functions compared to the
          $\tau$ spectral function as measured by ALEPH, CLEO and OPAL
          normalised to this $\tau$ spectral function. Figure taken from~\cite{ALEPHtau}.
          }
\end{minipage}
\end{figure}

The spectacular accuracy of 
$\alpha_{\mathrm{QED}}(0)=1/137.03599911(46)$
 measurement at low
$Q^2$ values~\cite{alphaqed0}
is spoiled when using it to predict the value at
higher mass scales, e.g. at the muon or Z mass.
In the correction that has to be made for the energy scale dependence:
$
\alpha_{\mathrm{QED}}^{-1}(Q^2)=\alpha_{\mathrm{QED}}^{-1}(0)
\left(
1-\Delta\alpha_{\mathrm{lep}}(Q^2)-\Delta\alpha_{\mathrm{had}}(Q^2)
\right)
$
the hadronic corrections, $\Delta\alpha_{\mathrm{had}}(Q^2)$,
are dominating the uncertainty.
At the energy scale corresponding to the Z mass this correction can
be expressed as~\cite{cabbibogatto}:\\[-2mm]
\[
\Delta\alpha_{\mathrm{had}}(M_{\mathrm{Z}}^2)=
-\frac{\alpha_{\mathrm{QED}}(0)\,s}{3\pi}
\int\limits_{s^{\prime}=4m_{\pi}^2}^{\infty}
\frac{R_{\mathrm{had}}(s^{\prime})}{s^{\prime}(s^{\prime}-M_{\mathrm{Z}}^2)}
\mathrm{d}s^{\prime}
+\Delta\alpha_{\mathrm{top}}(M_{\mathrm{Z}}^2),
\vspace*{-1.5mm}
\]
where $R_{\mathrm{had}}(s)$ is the ratio of the e$^+$e$^-$ hadronic
cross section over $\sigma($e$^+$e$^-$$\rightarrow\mu^{+}\mu^{-})$.
In calculating $\Delta\alpha_{\mathrm{had}}$,
the measured $R_{\mathrm{had}}(s)$ is used
in the five flavour limit for high energy. This is the reason an additional
correction for the top contribution has to be made, but this correction
is small and theoretically well under control.
The most recent result using this technique to estimate the hadronic
corrections, and the one preferred by
the LEP EW working group, by Burkhardt and Pietrzyk~\cite{burkhardt},
is shown in Fig.~\ref{alphaqed} and yields
$\Delta\alpha_{\mathrm{had}}^{(5)}=0.02758(35)$.
It is improved with respect to earlier estimates by using
new KLOE~\cite{KLOEalphaqed} data taken at DA$\Phi$NE
and CMD-2~\cite{CMD2alphaqed}, SND~\cite{SNDalphaqed} data taken at VEPP
below 1.1 GeV. The uncertainty for the QED
coupling constant at the mass of the Z is now dominated by the knowledge
of the e$^+$e$^-$ annihilation process into hadrons in the $\sqrt s$
range from 1.1 to 5 GeV.

The ALEPH collaboration has supplied a paper
summarising all its $\tau$ results, including a complete list of
exclusive decays and spectral functions~\cite{ALEPHtau}.
The Belle collaboration also showed $\tau$ results in the parallel session,
including a spectral function and a new m$_{\tau}$ measurement~\cite{BELLEtau}.
The $\tau$ spectral functions provide another approach to estimate the
hadronic corrections to the running of $\alpha_{QED}$.
A comparison of e$^+$e$^-$ and $\tau$ spectral functions is given in
Fig.~\ref{specfun}~\cite{ALEPHtau}.
This figure shows that there is a discrepancy between the e$^+$e$^-$ and
$\tau$ spectral functions, which may or may not be due to systematic effects
in the theory needed to translate the experimental data into spectral
functions. Note however that also the e$^+$e$^-$ data from the different
experiments are not fully consistent. The uncertainty that is normally assigned
to $\alpha_{\mathrm{QED}}$ and the anomalous magnetic moment of the
muon, mentioned below, is probably underestimated.

The hadronic correction to the electromagnetic coupling constant is also
an important ingredient to the uncertainty on the prediction of the muon
anomalous magnetic moment.
Translating the knowledge on $\alpha_{\mathrm{QED}}$ into a prediction for
$a_{\mu}=(g-2)_{\mu}/2$ we see that the predicted value differs
by 0.7 to 2.8 standard deviations from $a_{\mu}=11659214(9)\times 10^{-10}$,
the value measured by the
Muon $(g-2)$ Collaboration~\cite{BNLE821} (see Fig.~\ref{amu}.)
Clearly more work and data is needed to clear up this situation
further.\\
For the moment there is no reason to think that the SM description is not
in correspondence with the data in the QED sector.

\section{Weak Interactions}
\begin{figure}[bp]
\vspace*{-5mm}
\begin{minipage}[b]{7cm}
\centerline{
\includegraphics[height=7.3cm]{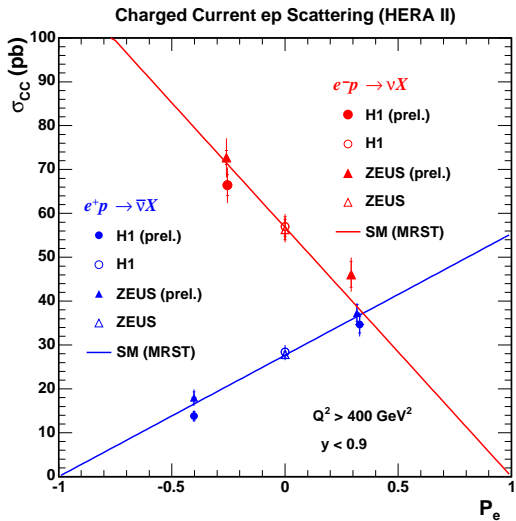}}
\caption[CC cross section versus polarisation]{\label{CCcs}
          The total charged current cross section for electron-proton and
          positron-proton scattering at $\sqrt{s}=318$~GeV as a function
          of polarisation of the lepton beam as measured by the
          H1 and ZEUS collaborations~\cite{H1ZEUSCCcs}.
         }
\end{minipage}
\hfill
\begin{minipage}[b]{7.5cm}
For right handed electrons (left handed positrons)
the CC cross section becomes zero while it increases linearly with
polarisation as expected for a pure $V-A$ coupling.

The H1 collaboration also made a fit to the vector and axial-vector
coefficients of the coupling, which is shown in Fig.~\ref{H1CDFud}.
Although the measurements from LEP~\cite{LEP1final} and CDF~\cite{CDFewfit}
are more precise, only the H1 result~\cite{H1ewfit} is able to
distinguish clearly 
between the two sign ambiguities
for both the u and d quark case.

In the first phase of LEP (LEP-1) and at SLD, where data were collected at CM
energies near the Z mass, the couplings of the Z boson to fermions were
investigated in detail.

The analysis of these data is in its final stages
(The final LEP-1 results have been published after this conference
in~\cite{LEP1final})
and most results were not
updated for this conference, except for the heavy flavour
results that are now all final.
\end{minipage}
\end{figure}

\vspace*{-3mm}
To study the weak interaction the H1 and ZEUS experiments,
at the HERA collider have collected some 200 pb$^{-1}$ of polarized e$^+$
and e$^-$ collision data on protons at $\sqrt s=318$ GeV.
The total Charged Current (CC) cross section
($\mathrm{ep}\rightarrow \nu\mathrm{X}$) versus $\mathrm{e}^{\pm}$ polarisation
plotted in Fig.~\ref{CCcs} shows
that the exchange of a charged W boson results in a purely
vector minus axial-vector coupling.

\begin{figure}[tbp]
\centerline{
\includegraphics[height=7cm]{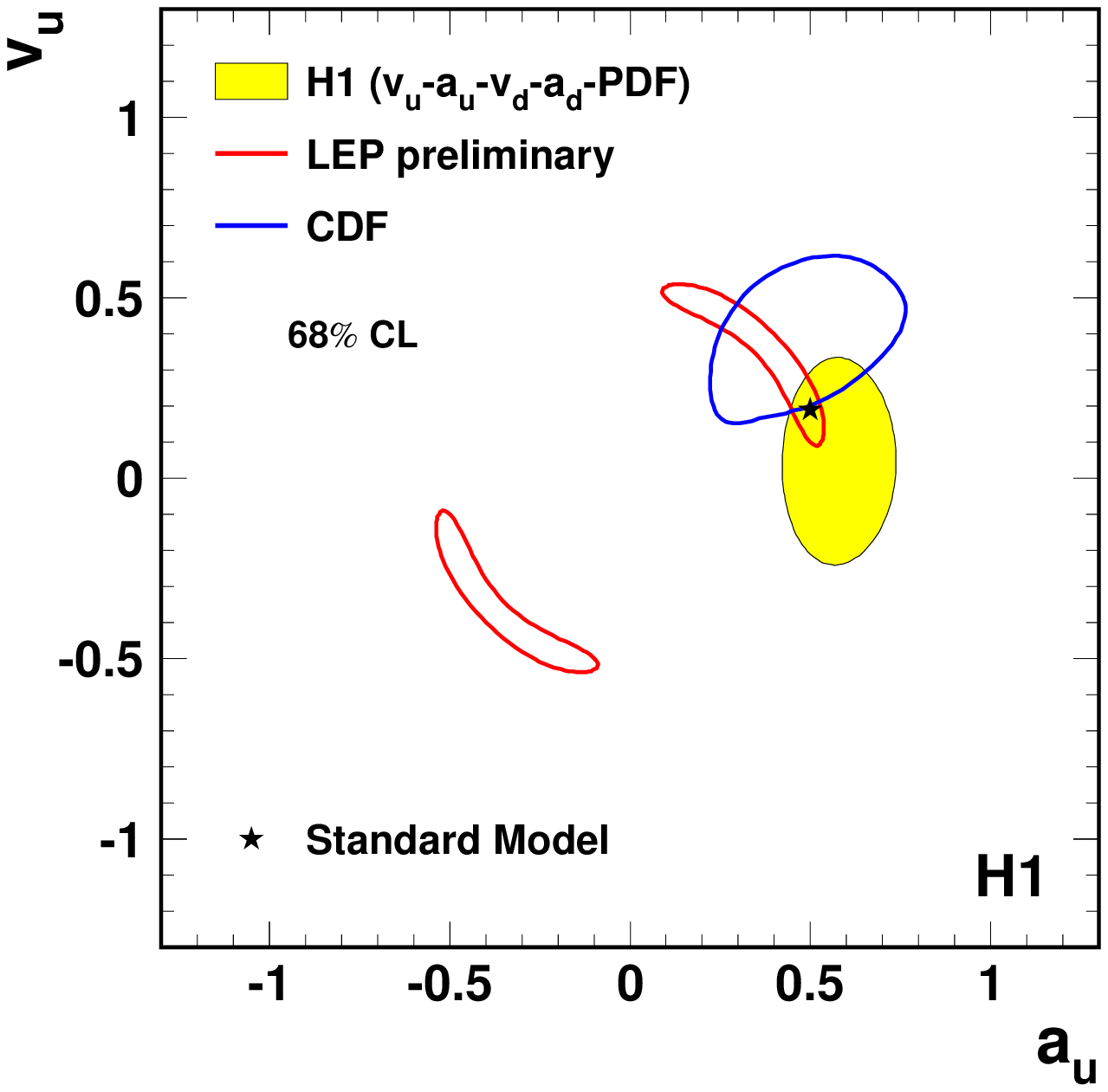}\hfill
\includegraphics[height=7cm]{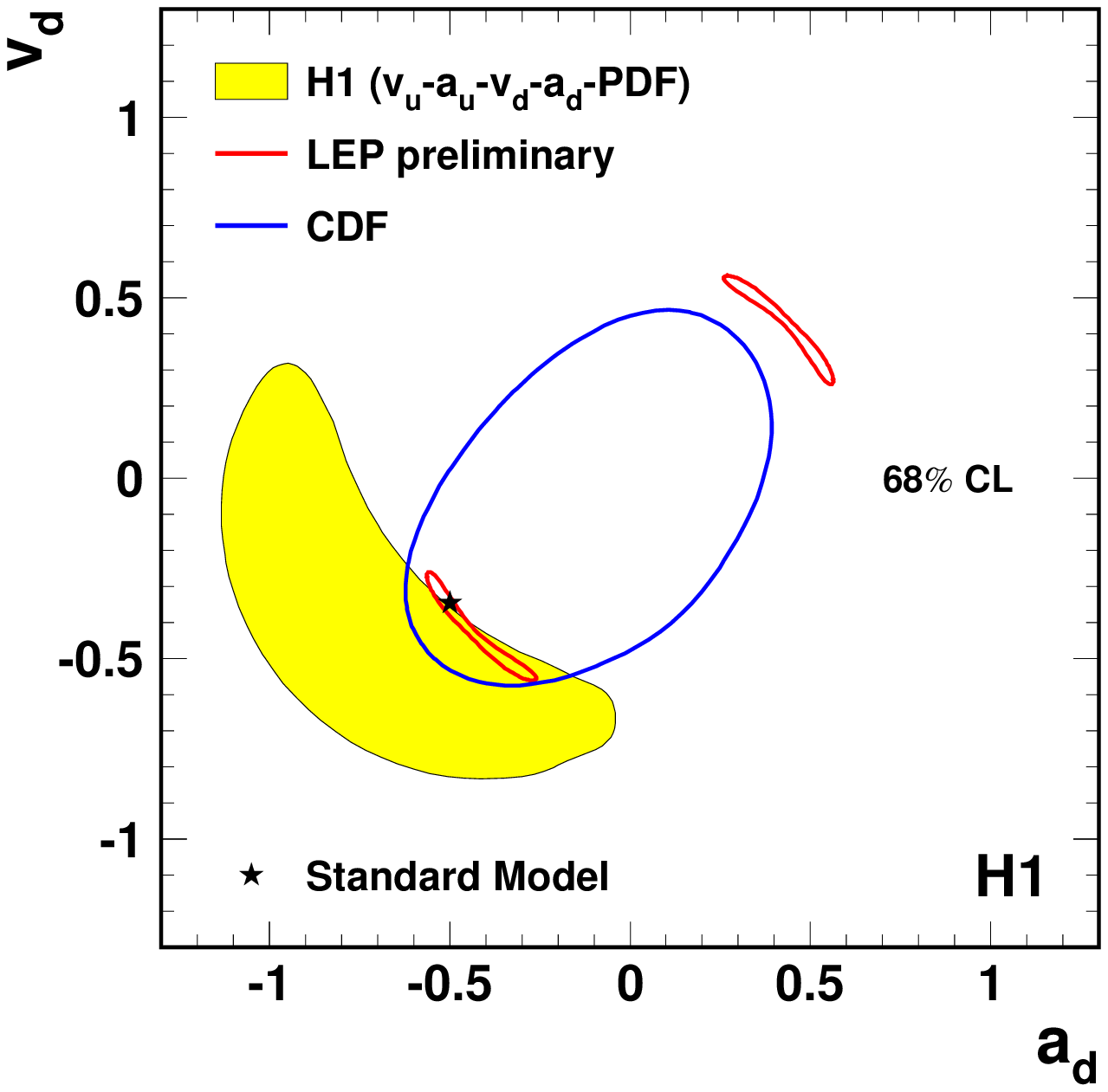}}
\caption[Up and down couplings from H1]{\label{H1CDFud}
          Vector and axial-vector coupling coefficients for u quarks (left)
          and d quarks (right) as measured by
          H1~\cite{H1ewfit}, CDF~\cite{CDFewfit}
          and the LEP experiments~\cite{LEP1final}.
         }
\end{figure}

The heavy flavour measurements from the LEP experiments and SLD
used in this heavy flavour fit are:
\begin{itemize}
\item \vspace*{-2mm}
 $R_q=\frac{\sigma(\mathrm{e}^+\mathrm{e}^-\rightarrow\;
                                             q\overline{q})}{
                                             \sigma(\mathrm{e}^+\mathrm{e}^-\rightarrow\;
                                             \mathrm{hadrons})},\;q=\mathrm{c},\mathrm{b}$,
the partial hadronic widths into b and c quarks.
The experimental issue is the clean identification of c and b quarks,
while the theoretical challenge is to correct for higher order production of
heavy quark pairs.
\item \vspace*{-2mm}
 $A_{\mathrm{FB}}^{0,q}=\frac{\sigma_{\mathrm{F}}-\sigma_{\mathrm{B}}}{
\sigma_{\mathrm{F}}+\sigma_{\mathrm{B}}}=A_{\mathrm{e}} A_{q},
\;q=\mathrm{c},\mathrm{b}$,
the heavy quark forward-backward asymmetries, where the first part
of the formula indicates the experimental method, to distinguish
the direction of the heavy quark and anti-quark. The direction is called forward (F)
if the quark moves in the direction of the incoming electron and
backward (B) in the other situation.
Theoretically this quantity can be expressed in the
product of an asymmetry generated at the Zee, $A_{\mathrm{e}}$ vertex
and at the Z$q\overline{q}$ ($q=$c,b) vertex, $A_q$.
\end{itemize}
\vspace*{-2mm}
At SLD also polarised beams are used, which make the measurement of
the asymmetries using left- and right-handed polarised beams possible:
\begin{itemize}
\item \vspace*{-2mm}
$A_{\mathrm{LR}}^{0}=
\frac{\sigma_{\mathrm{L}}-\sigma_{\mathrm{R}}}{
\sigma_{\mathrm{L}}-\sigma_{\mathrm{R}}}
A_{\mathrm{e}}=
\frac{2\,v_{\mathrm{e}}\,a_{\mathrm{e}}}{v_{\mathrm{e}}^2+a_{\mathrm{e}}^2}$,
the left-right asymmetry for the total cross section,
where also the relation of this quantity and the coupling coefficients is given.
In addition to the considerations given above for heavy flavour measurements,
experimentally knowledge of the degree of polarisation of the beams is crucial
here. Theoretically the measurement is very clean and many systematic effects,
both experimental and theoretical, cancel in asymmetry measurement.
\item \vspace*{-2mm}
$A_{\mathrm{LRFB}}^{0,q}=A_{q}=
\frac{2\,v_q\,a_q}{v_q^2+a_q^2},
\;(q=\mathrm{c},\mathrm{b})$,
the combined forward-backward--left-right asymmetry for the total cross section,
where also the relation of this quantity and the coupling coefficients is given.
The same considerations as in the previous item apply.
\end{itemize}
\vspace*{-2mm}
\begin{figure}[tp]
\vspace*{-1cm}
\begin{minipage}[t]{7.25cm}
\centerline{\includegraphics[height=6cm]{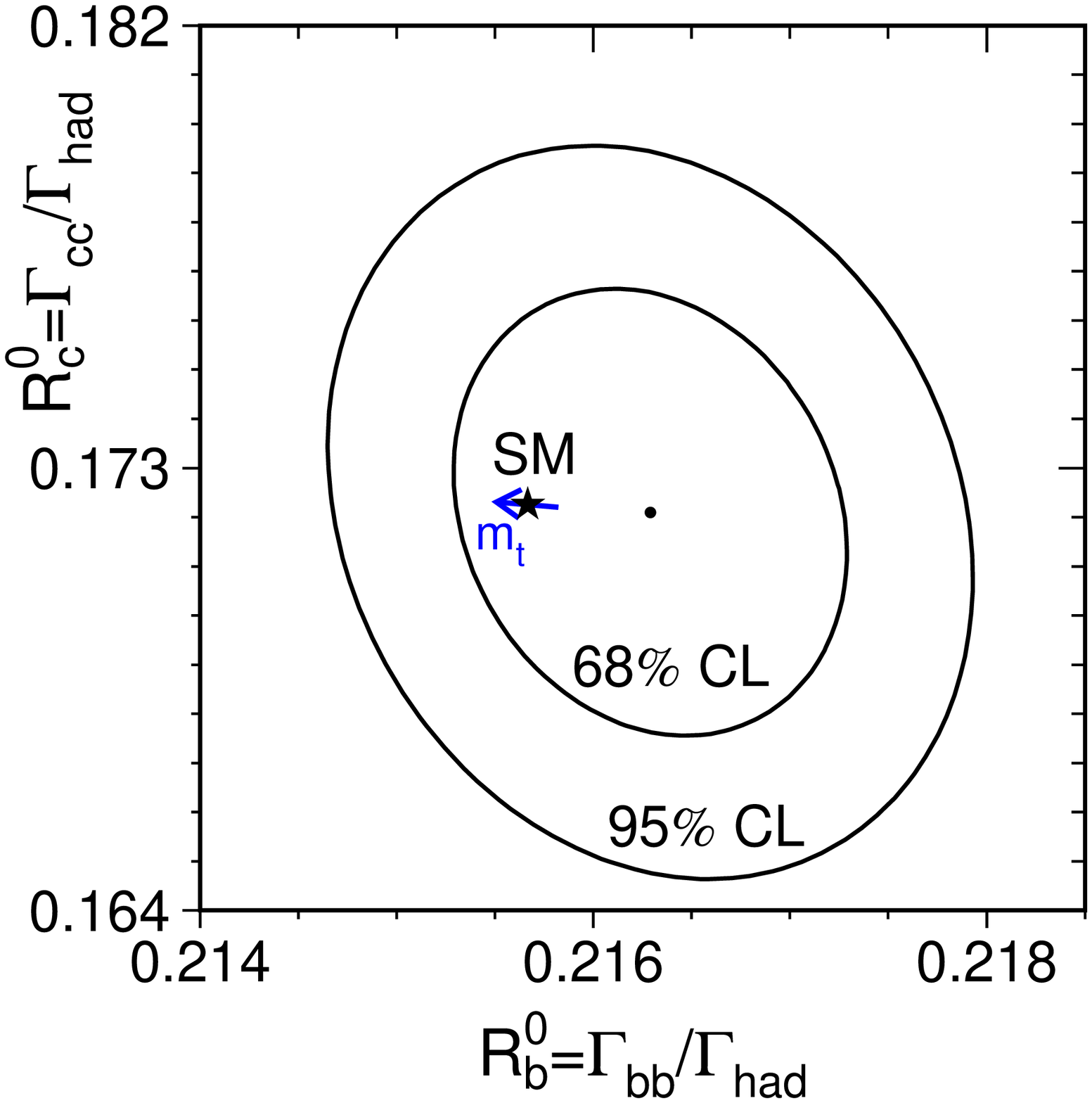}}
\caption[Rc versus Rb]{\label{RcRb}
          $R_{\mathrm{b}}$ versus $R_{\mathrm{c}}$ 
          from the final LEP EW working group fit to the LEP and SLD
          heavy flavour data.
          }
\end{minipage}
\hfill
\begin{minipage}[t]{7.25cm}
\centerline{\includegraphics[height=6cm]{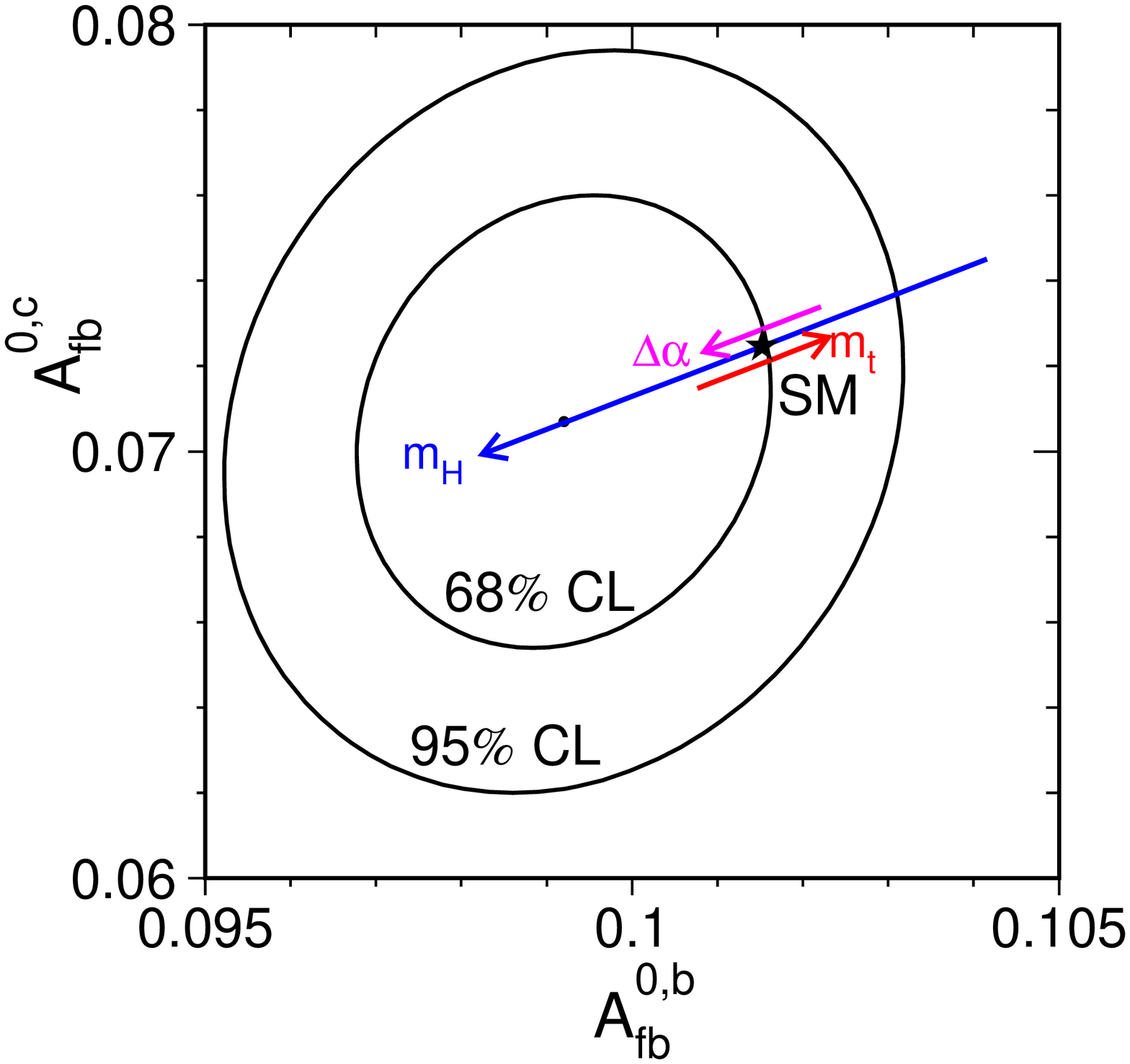}}
\caption[Afbc versus Afbb]{
          \label{AfbcAfbb}
          $A_{\mathrm{FB}}^{0,\mathrm{c}}$ versus $A_{\mathrm{FB}}^{0,\mathrm{b}}$
          from the LEP EW working group fit to the LEP and SLD
          heavy flavour data.
                    }
\end{minipage}
\vspace*{-3mm}
\end{figure}
\begin{figure}[bp]
\vspace*{-4mm}
\begin{minipage}[t]{7.25cm}
\centerline{
\includegraphics[height=8.7cm]{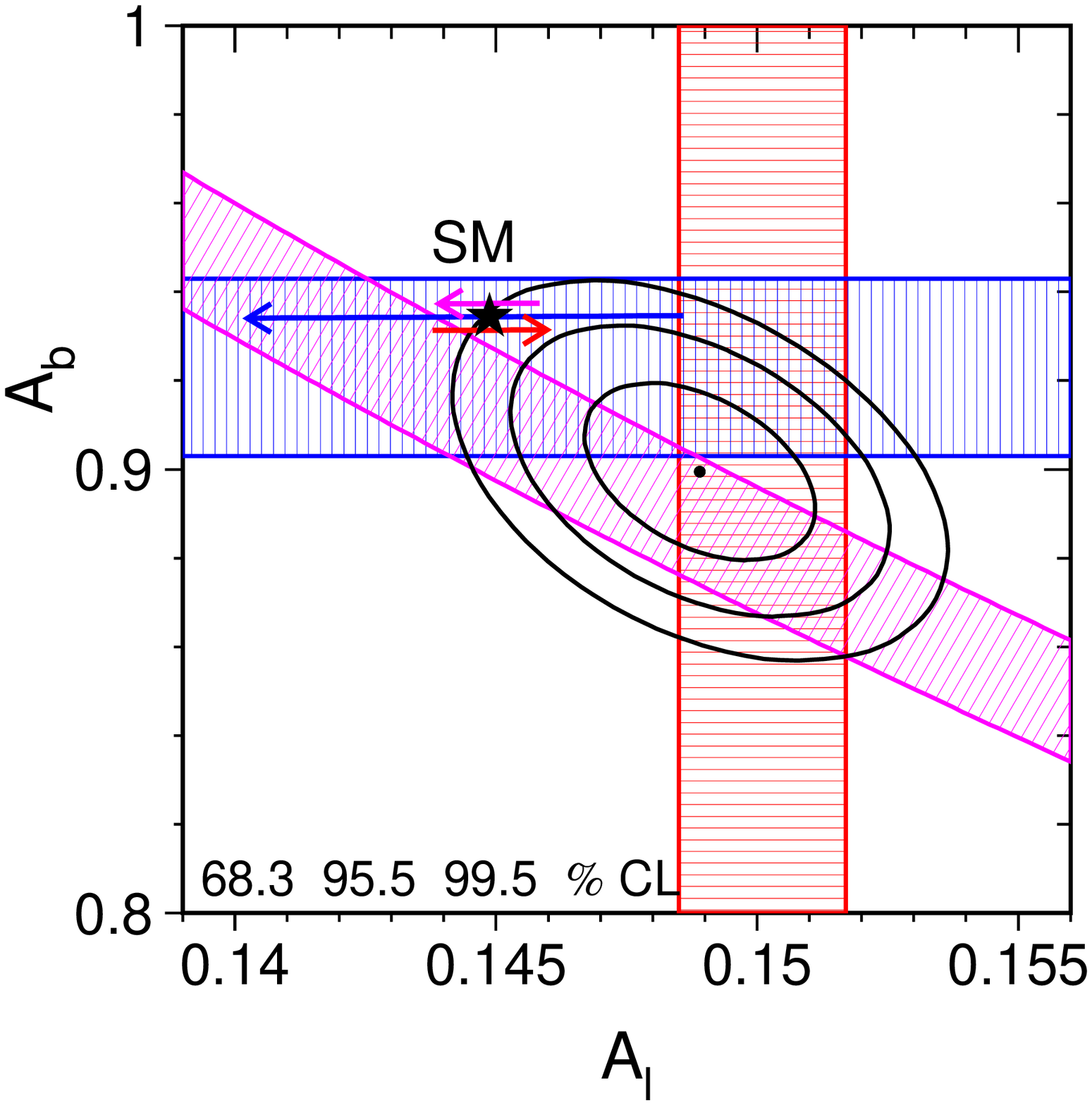}}
\caption[Al versus Ab]{
              \label{AlAb}
              The lepton asymmetry versus b-quark asymmetry.
              The horizontal and vertical bands are from SLD measurements for
              $A_{\mathrm{e}}$ and $A_{\mathrm{b}}$. The
              anti-diagonal is the LEP $A_{\mathrm{FB}}^{(0,b)}$
              measurement.
              The SM prediction is the star, with the 
              influence of $1\sigma$ varying $\Delta\alpha_{\mathrm{had}}$,
              $m_{\mathrm{H}}$ and $m_{\mathrm{t}}$ indicated by the arrows
              in top to bottom order.
              }
\end{minipage}
\hfill
\begin{minipage}[t]{7.25cm}
\centerline{
\includegraphics[height=8.7cm]{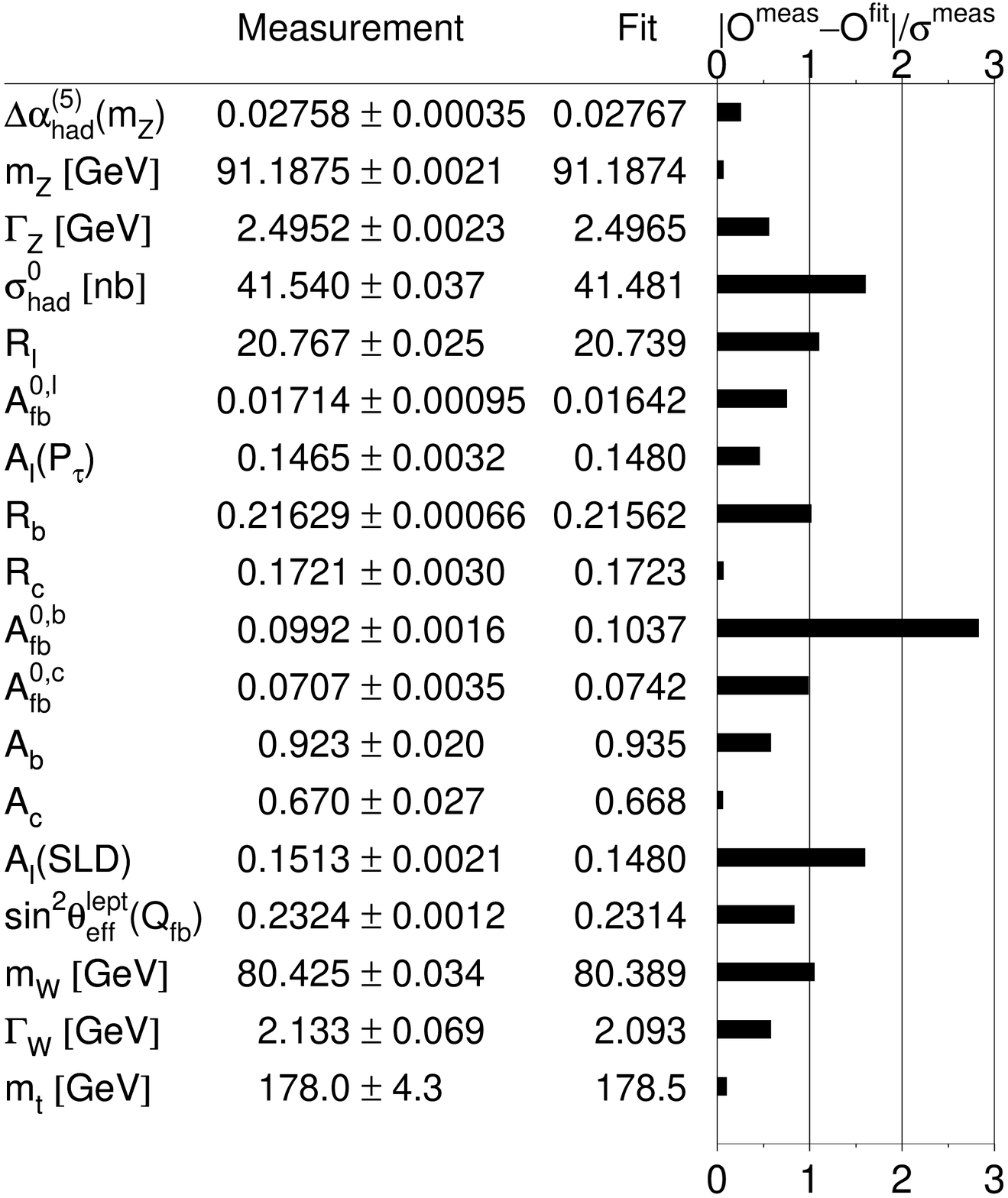}}
\caption[LEP/SLD EW measurements]{
              \label{LEPEWtable}
              Electroweak parameter measurements as derived by the LEP
              EW working group on the basis of ALEPH, DELPHI, L3, OPAL,
              SLD, CDF and D\O\ measurements.
              }
\end{minipage}
\end{figure}
Another way to determine the lepton asymmetry is by measuring the $\tau$
polarisation asymmetry $A_{\tau}$, which is done in $\tau$ decays at LEP.
In the following the SM relation $A_{\mathrm{e}}=A_{\tau}$ is used.

Combining the heavy flavour measurements from the LEP experiments and
SLD and fitting the relevant SM parameters to these measurements
a goodness of fit of $\chi^2$/d.o.f.$=53/91$ is obtained,
indicating excellent overall agreement.
This is illustrated in Fig.~\ref{RcRb} and~\ref{AfbcAfbb},
which show the measured probability contours for 
$R_{\mathrm{c}}$ versus $R_{\mathrm{b}}$ and
$A_{\mathrm{FB}}^{0,\mathrm{c}}$ versus $A_{\mathrm{FB}}^{0,\mathrm{b}}$
respectively with the SM predictions.
The arrows in these figures show the sensitivity to other SM parameters,
such as the mass of the top quark, the Higgs boson mass and the size
of the hadronic correction to the electromagnetic coupling strength.

Digging a little deeper into all possible comparisons between the measurements
and the SM predictions one comes across the situation as depicted in
Fig.~\ref{AlAb}.
Although the $A_{\ell}=A_{\mathrm{e}}$ and $A_{\mathrm{b}}$
measurements each agree fairly well with
the SM prediction and despite the fact that the measurements agree on a
unique $(A_{\mathrm{e}}, A_{\mathrm{b}})$ point, there is a discrepancy at
the $>3\sigma$ level with the SM prediction for this combination.
It should be noted that this is the biggest discrepancy and a single one
out of many possible comparisons of the SM to the data.

Apart from the heavy flavour results, the LEP experiments and SLD,
produced a number of other parameters at or around a center
of mass energy corresponding to the Z boson mass:
the mass and width of the Z boson, $m_{\mathrm{Z}}$ and
$\Gamma_{\mathrm{Z}}$, the cross section
of e$^+$e$^-$$\rightarrow$hadrons, $\sigma_{\mathrm{had}}^0$,
the ratio of the hadronic to muon-pair final state at the Z peak,
$R_{\ell}$, the forward-backward asymmetry for the lepton-pair final
state, $A_{\mathrm{FB}}^{0,\ell}$, and the effective lepton
weak mixing angle
as derived from the forward-backward asymmetry
for a Z decaying into quark-pairs using jet-charge techniques,
$\sin^2\theta_{\mathrm{eff}}^{\mathrm{lept}}(Q_{\mathrm{FB}})$.
From running LEP at CM energies above the W-pair threshold
also the mass and width of the W boson have been determined,
$m_{\mathrm{W}}$ and $\Gamma_{\mathrm{W}}$.
The present results as derived by the LEP electroweak working group
are listed in Fig.~\ref{LEPEWtable}.
The largest discrepancy in this table is the b quark forward backward
asymmetry, which was discussed before.

\begin{figure}[tbp]
\begin{minipage}[t]{9cm}
\centerline{
\includegraphics[height=6cm]{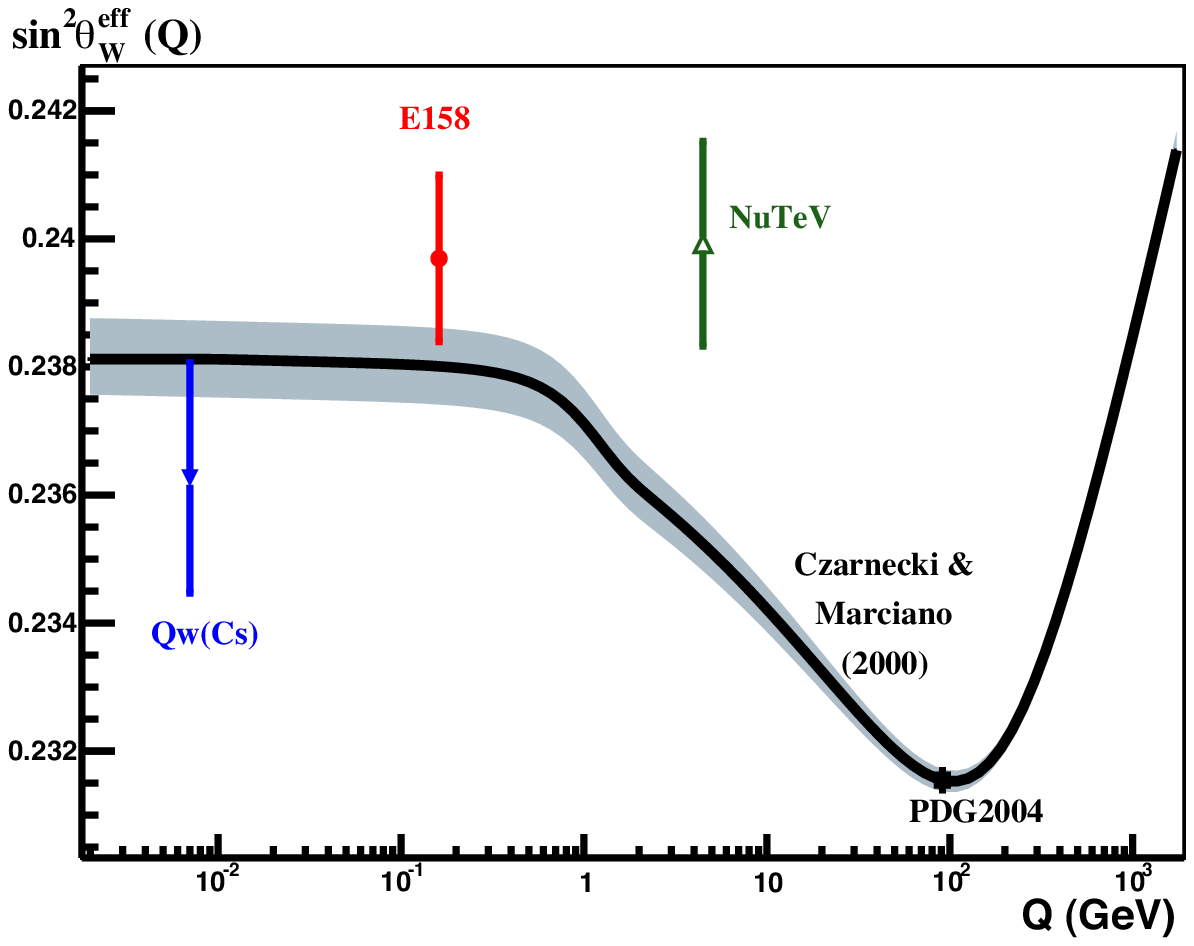}}
\caption[Running of sin2theta]{
              \label{sin2theta}
              The curve is $\sin^2\theta_w^{\mathrm{eff}}$ predicted by the SM~\cite{CandM},
              as a function of $Q=\sqrt{|Q^2|}$ when fixed to the value at
              $Q^2=M_{\mathrm{Z}}^2$ from~\cite{PDG}
              based on LEP and SLD measurements.
              The measurements at $Q\approx 0$ are from atomic parity violation,
              $Q_w$(Cs)~\cite{cesium} and
              from M{\o}ller scattering~\cite{E158}.
              At $Q$ values in the range of a few GeV is the neutrino-nucleon scattering
              result from NuTeV~\cite{nutevsin2theta}.
              }
\end{minipage}
\hfill
\begin{minipage}[t]{5.5cm}
\centerline{
\includegraphics[height=6cm]{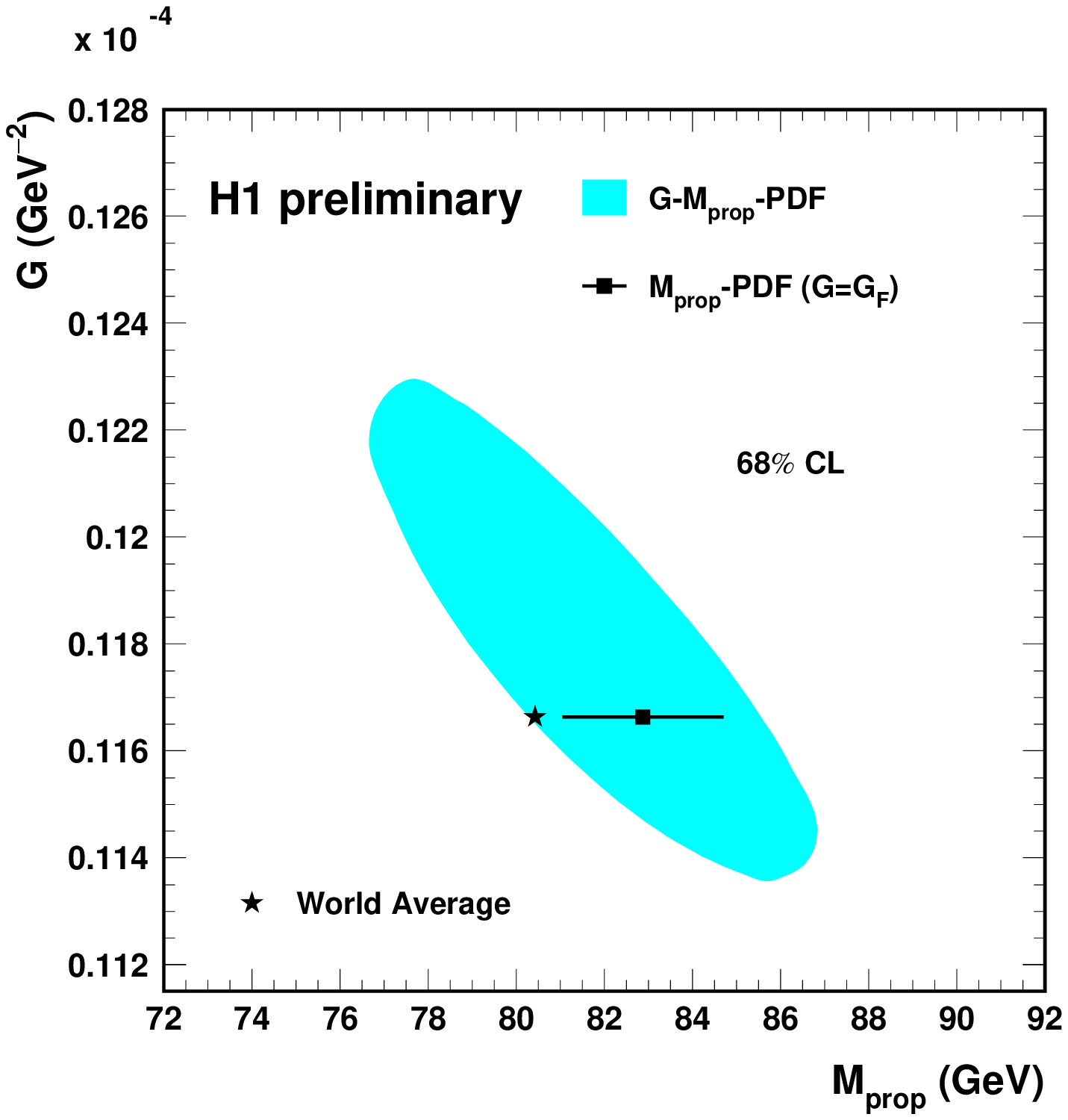}}
\caption[H1 GF versus mW]{
              \label{H1GFvsmW}
              The Fermi coupling strength, $G_F$ versus the mass of the W boson
              as measured by the H1 collaboration~\cite{H1ewfit}.
              }
\end{minipage}
\vspace*{-4mm}
\end{figure}
Confronting the results from LEP, SLD and Tevatron, cast in the form
of sin$^2\theta_w$ in a $Q^2$ dependent extrapolation by
Czarnecki and Marciano~\cite{CandM},
with lower energy measurements
we see in Fig.~\ref{sin2theta}
good agreement with the atomic parity violation experiment
on Cesium~\cite{cesium}
and low energy M{\o}ller scattering from the E158 experiment~\cite{E158}.
For the NuTeV measurement~\cite{nutevsin2theta}
the situation has not changed since last year.
It does not fit well with the expectation in the $Q^2$ range
where the measurement has been made, being nearly three standard
deviations off. It must be noted that in neutrino-nucleon scattering the
$Q^2<0$, while in this figure it is compared to a determination at
LEP at $Q^2>0$. The exact $Q^2$ in the NuTeV experiment varies event
by event and this is accounted for by effectively fitting a running
$\sin^2\theta$ to the data. The effect of the uncertainty on the average
$Q^2$ scale is determined to be very small in~\cite{ZellerThesis}.

\section{The electroweak vector boson properties}
\vspace*{-3mm}
The H1 collaboration delivered a first EW fit of their data, being able to
extract the Fermi coupling strength and the mass of the W boson
simultaneously~\cite{H1ewfit}, as shown in Fig.~\ref{H1GFvsmW}.

\begin{figure}[tp]
\begin{minipage}[t]{4.8cm}
\centerline{\includegraphics[height=5cm]{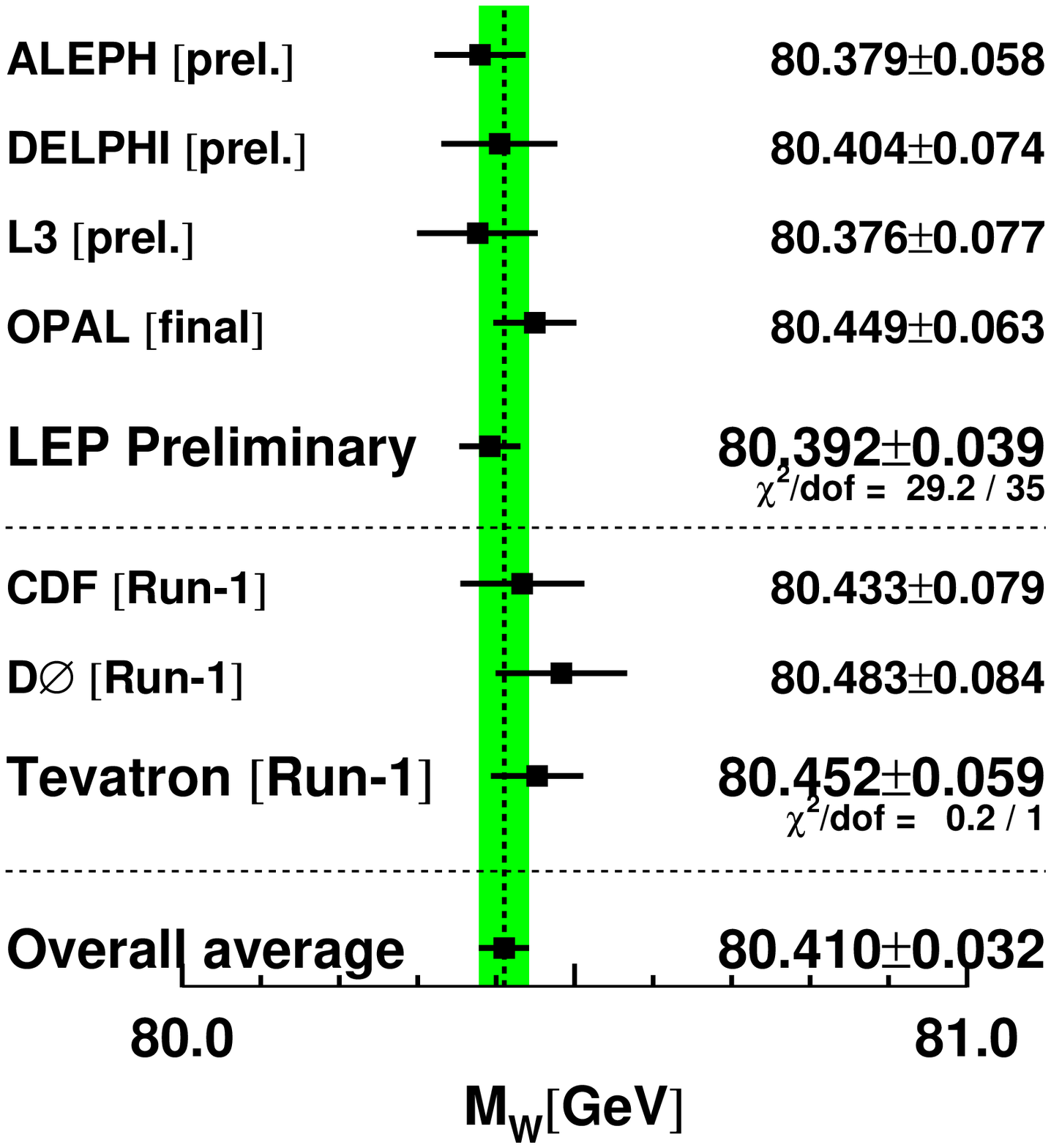}}
\caption[mW world average]{\label{Wmass}
                Summary and averages of measurements of $m_{\mathrm{W}}$.
                }
\end{minipage}
\hfill
\begin{minipage}[t]{4.8cm}
\centerline{\includegraphics[height=5cm]{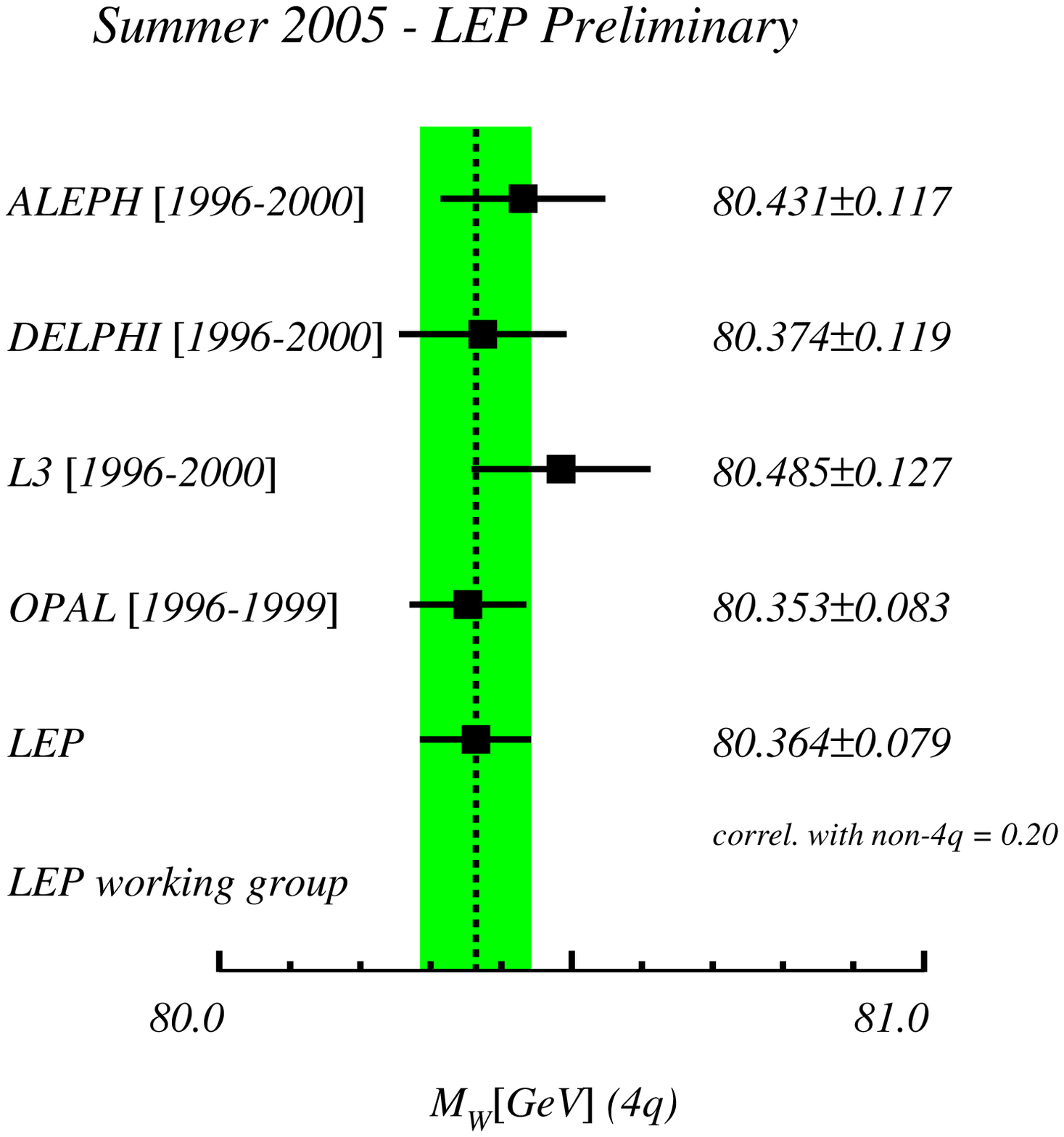}}
\caption[mW in the 4 jet channel]{\label{mWqqqq}
                Measurements of $m_{\mathrm{W}}$ in the 4-jet final state.
                 }
\end{minipage}
\hfill
\begin{minipage}[t]{4.8cm}
\centerline{\includegraphics[height=5cm]{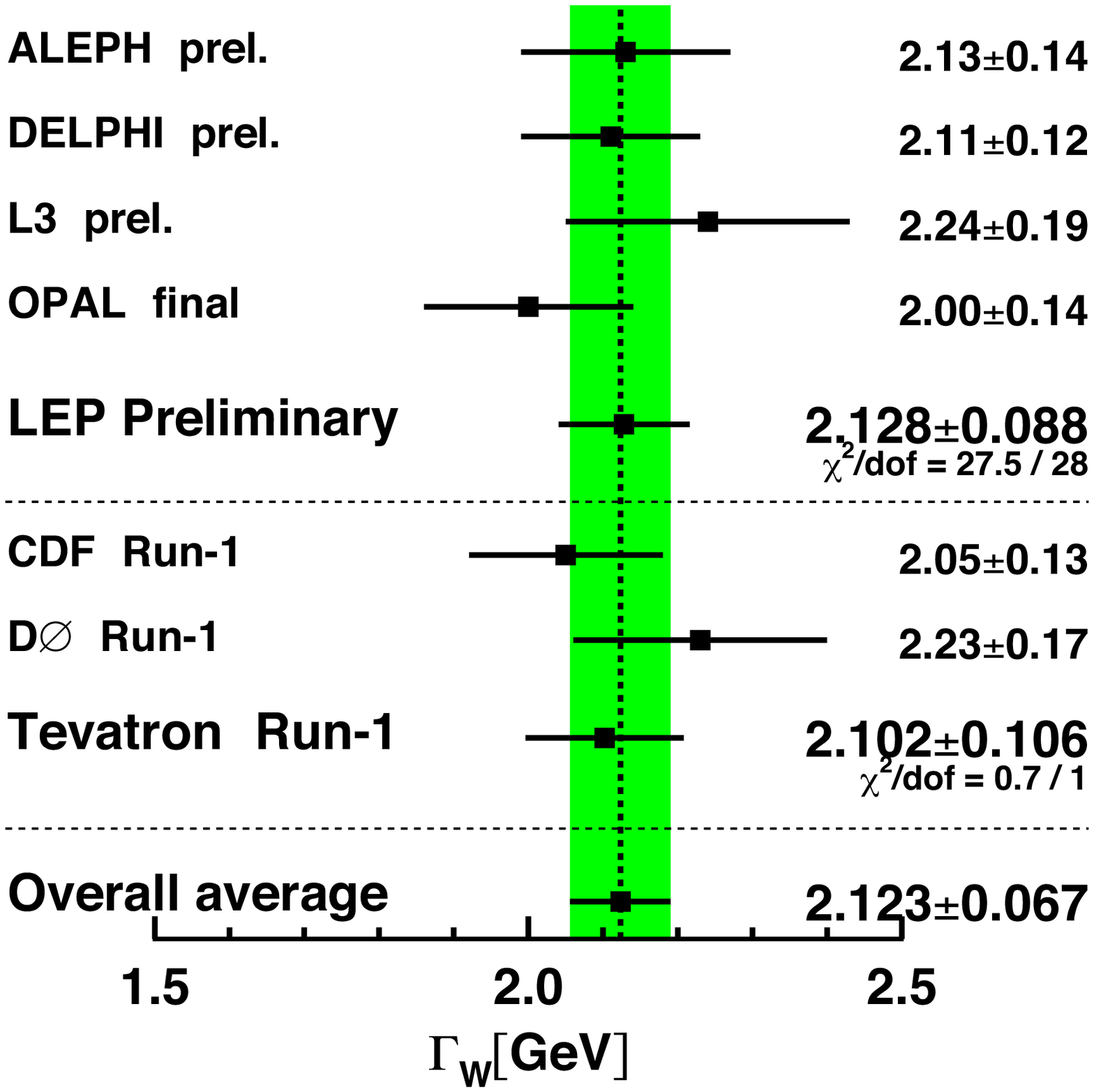}}
\caption[W width]{\label{GW}
                 Measurements and average of $\Gamma_{\mathrm{W}}$. 
          }
\end{minipage}
\vspace*{-3mm}
\end{figure}

The W boson mass has been measured precisely
at the second phase of LEP.
The results for the W boson mass measurement are
shown in Fig.~\ref{Wmass}.
The OPAL experiment produced final results for this conference,
while the ALEPH, DELPHI and L3 results are still preliminary.

\begin{figure}[bp]
\vspace*{-5.5mm}
\begin{minipage}[b]{9cm}
\centerline{\includegraphics[width=9cm]{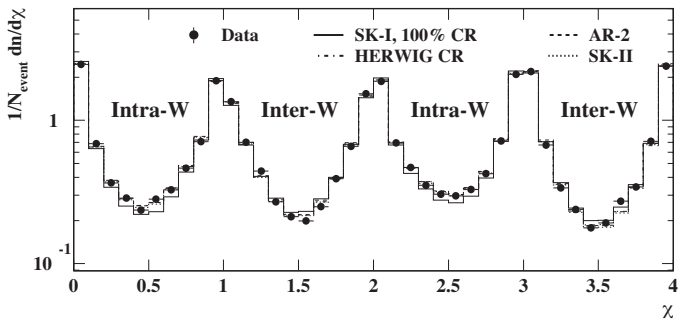}}
\caption[CR as measured in OPAL]{\label{CR}
                 Measurement of the intra- and inter-W particle densities by
                 the OPAL experiment~\cite{OPALCR}.
                 The measurement is compared to several Monte Carlo models
                 with and without colour reconnection.
          }
          \vspace*{3mm}
\end{minipage}
\hfill
\begin{minipage}[b]{5.3cm}
Sj\"{o}strand-Khoze-I model~\cite{SK1}
is estimated at 95\% CL.
Soft particles are more susceptible to
colour reconnection and Bose-Einstein
correlation effects. OPAL reduced
the colour reconnection uncertainty
by cutting on particle momentum.
The remaining uncertainty can be better estimated using
the method of the intra- and inter-W particle flow measurement to be
49~MeV for colour reconnection and 22~MeV for Bose-Einstein
correlations.
\end{minipage}
\vspace{-6mm}
\end{figure}

The final OPAL measurement introduces a number of refinements,
notably for the determination of possible systematic errors due to
colour reconnection between the quarks from different W boson decays in
WW$\rightarrow$4~jet events.
The particle flow is studied between jets from the same and from different W's
in the event. 
A comparison for these intra- and inter-W particle flows is made to different
models for and strengths of colour reconnection in Fig.~\ref{CR}.
The comparison reveals no significant sign of
colour reconnection and an upper limit of
37\% of the strength predicted by the

\noindent
These two effects still remain the leading contributions to
the overall systematic
uncertainty on the $m_{\mathrm{W}}$ measurement.
The current state for the $m_{\mathrm{W}}$ measurements from the
WW$\rightarrow$4~jet channel is shown in Fig.~\ref{mWqqqq}.
Similar reductions in error as for OPAL can be expected
for the final results for $m_{\mathrm{W}}$ in this channel
from the other LEP experiments.

The Tevatron experiments CDF and D\O\ had collected nearly
1~fb$^{-1}$ when this conference took place. Of that data volume
typically about 300~pb$^{-1}$ has been analysed for final and
preliminary results.
Potentially, the CDF and D\O\ experiments will be able
to measure the W boson mass with a precision similar to the
LEP results. Crucial ingredient in this measurement is the Jet Energy Scale,
which at present is not yet sufficiently under control to produce a
competitive measurement.

The final OPAL results also lead to a new average
$\Gamma_{\mathrm{W}}=2.123\pm 0.067$~GeV.
Figure~\ref{GW} shows the current state of the $\Gamma_{\mathrm{W}}$
measurements and average value.

\begin{figure}[tp]
\vspace*{-5mm}
\begin{minipage}[t]{7.25cm}
\centerline{\includegraphics[height=6cm]{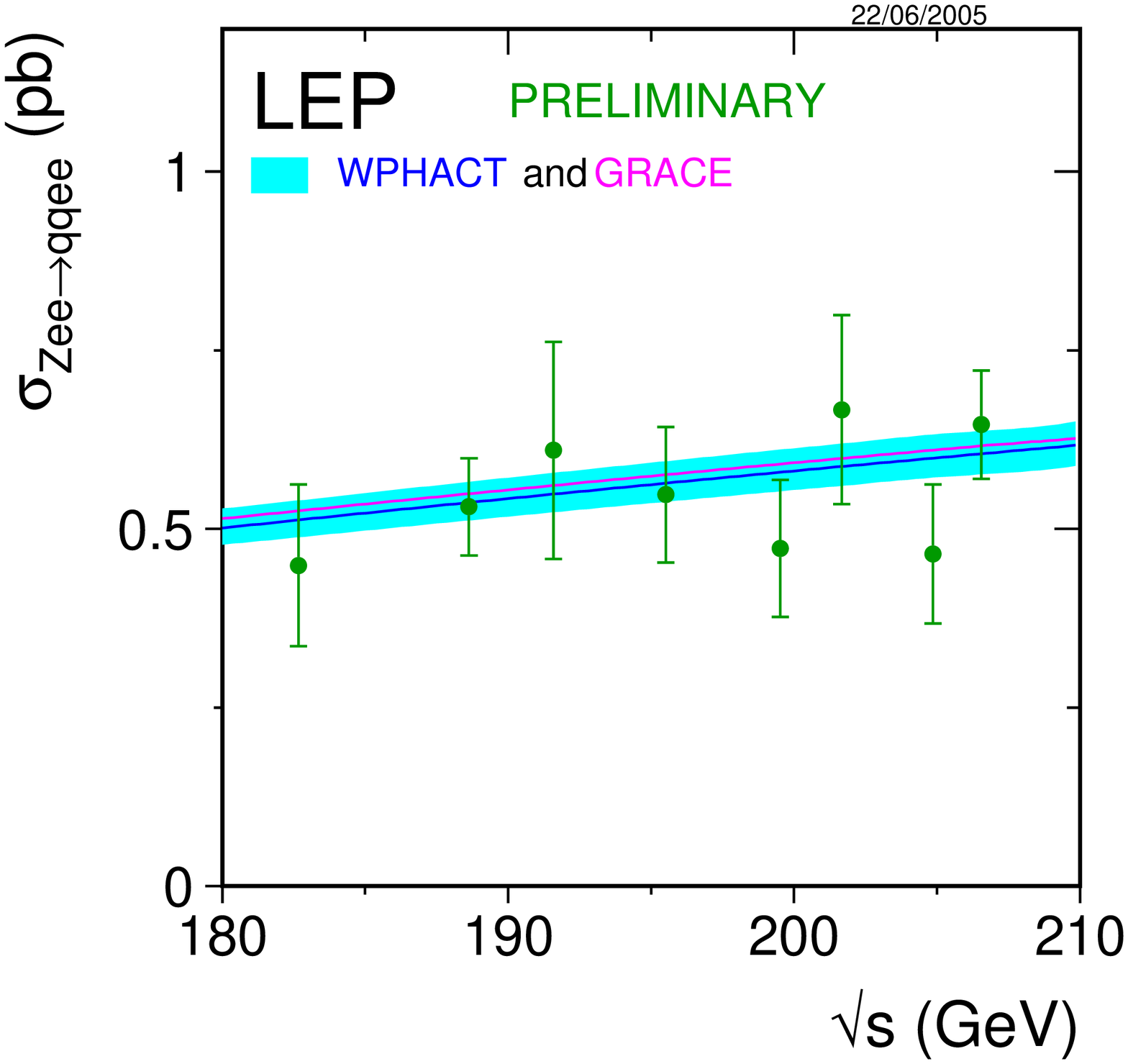}}
\end{minipage}
\hfill
\begin{minipage}[t]{7.25cm}
\centerline{\includegraphics[height=6cm]{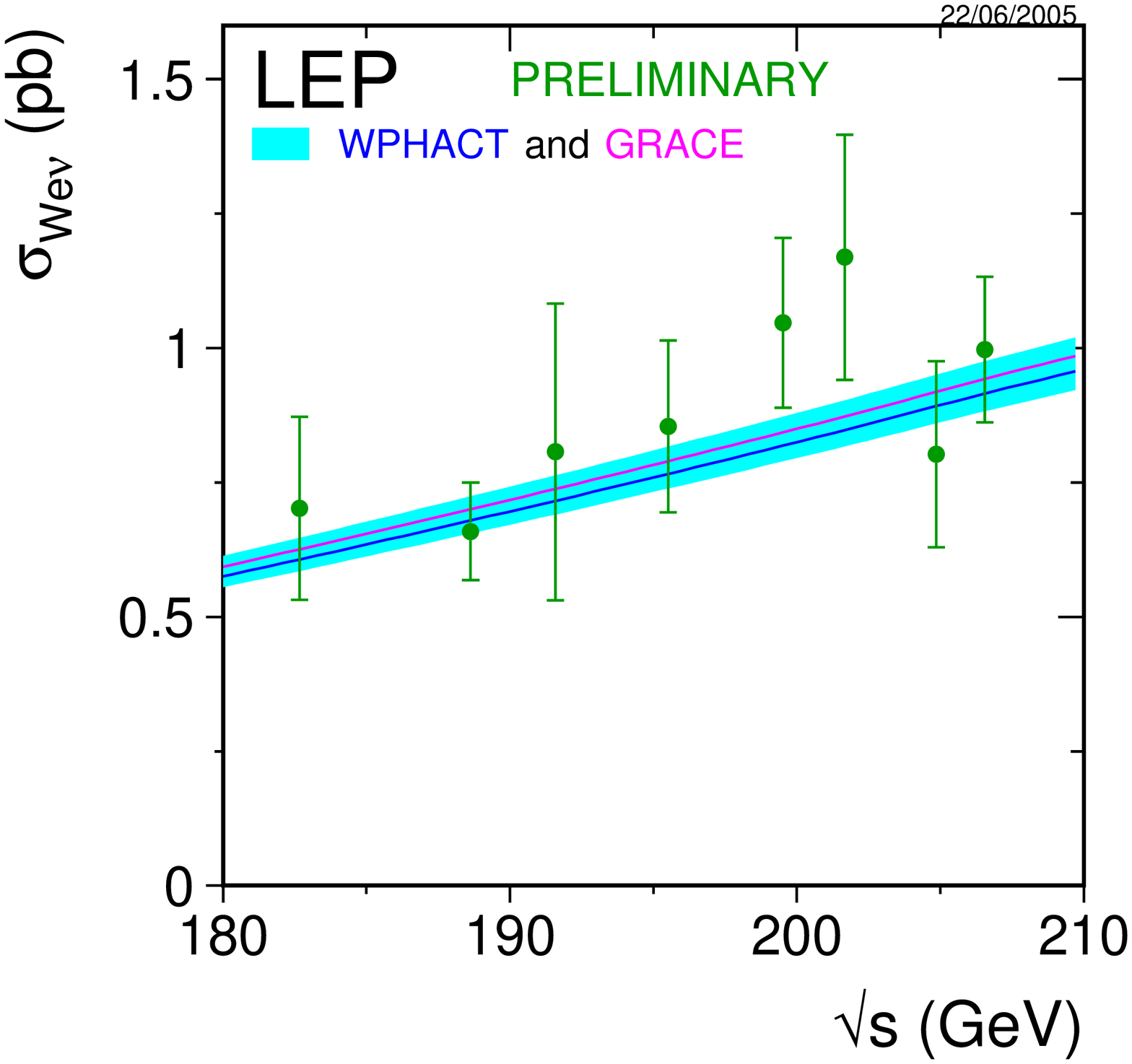}}
\end{minipage}
\vspace*{-3mm}
\caption[single Z and W production at LEP]{\label{singleWZ}
                Cross section measurements for single Z (left) and W (right)
                boson production at LEP for energies between 180 and 210~GeV.
                Predictions based on WPHACT and Grace are also plotted.
                    }
                    \vspace*{-4mm}
\end{figure}

At LEP-2 Z and W bosons can also be singly produced in a
Zee and We$\nu$ final state respectively, which is well described
by the SM as is shown in Fig.~\ref{singleWZ}.

\begin{figure}[bp]
\vspace*{-3mm}
\begin{minipage}[t]{7.25cm}
\centerline{\includegraphics[height=6cm]{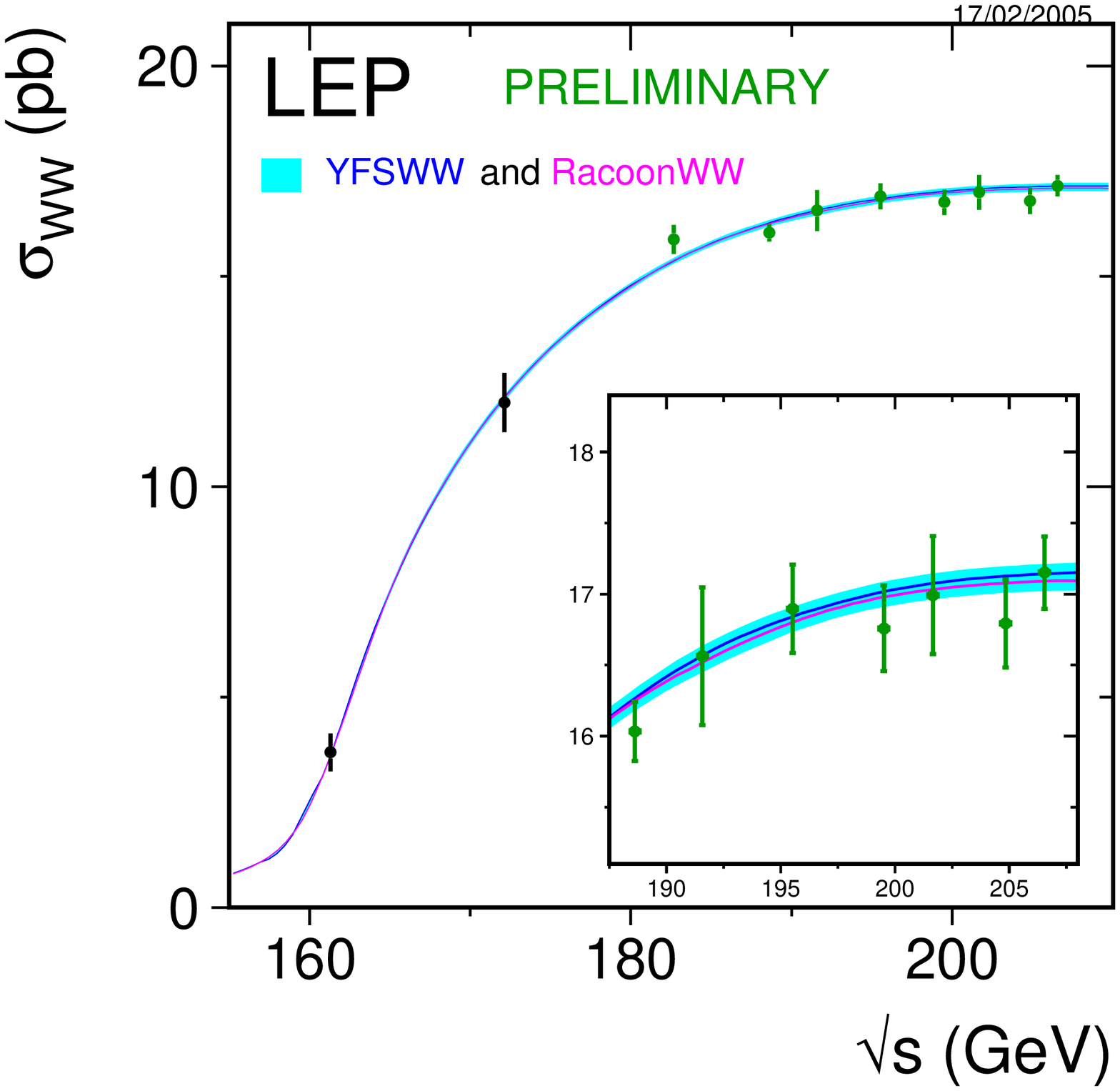}}
\caption[W-pair cross section at LEP]{\label{sigmaWW}
                Cross section measurements for W-pair production at LEP
                and the predictions by the SM (YFSWW/RacoonWW).
                    }
\end{minipage}
\hfill
\begin{minipage}[t]{7.25cm}
\centerline{\includegraphics[height=6cm]{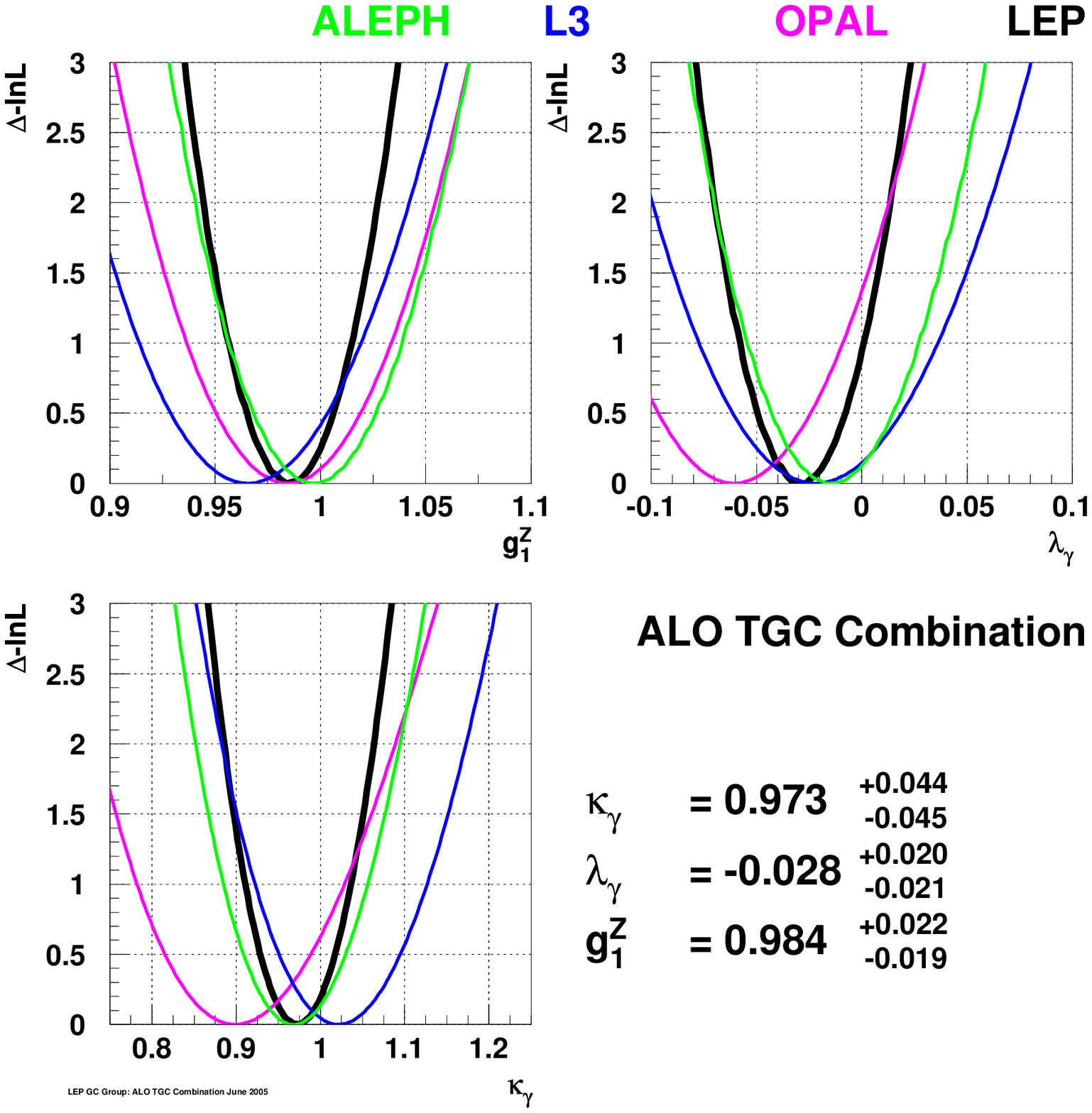}}
\caption[LEP Triple Gauge Boson Couplings]{\label{LEPTGC}
                Fits for the W boson anomalous triple gauge boson
                couplings~\cite{LEPTGC},
                based on W-pair production cross section and angular
                distributions as measured by ALEPH~\cite{ALEPHTGC},
                L3~\cite{L3TGC} and OPAL~\cite{OPALTGC}.
          }
\end{minipage}
\vspace*{-3mm}
\end{figure}
\begin{figure}[tbp]
\begin{minipage}[t]{7.25cm}
\centerline{\includegraphics[height=7.5cm]{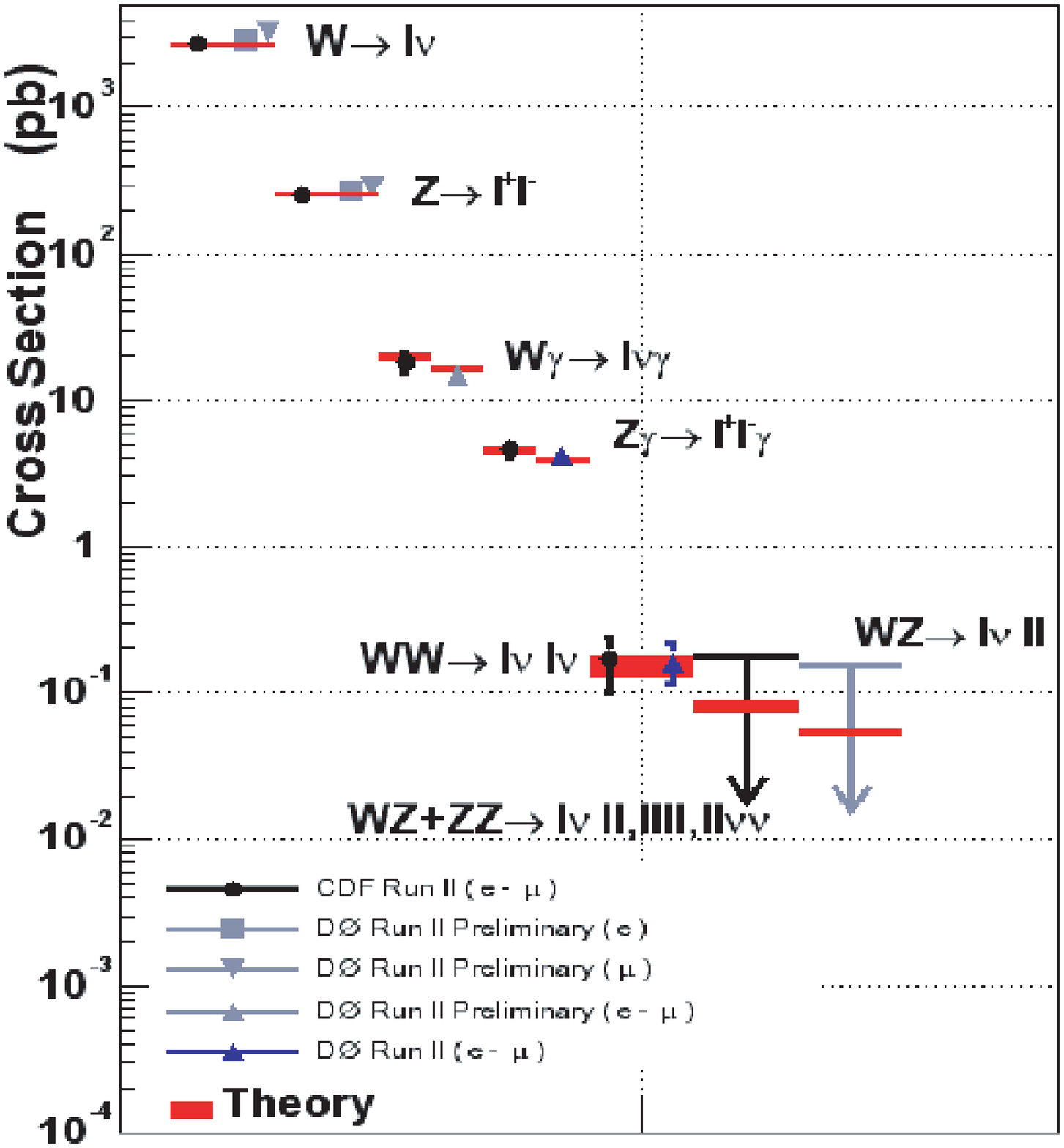}}
\caption[Multi boson production at the Tevatron]{\label{TeVmultiboson}
                Cross sections for the production of multi boson final states
                at the Tevatron. The lines with an arrow downwards give
                measured upper limits for the cross section.
                    }
\end{minipage}
\hfill
\begin{minipage}[t]{7.25cm}
\centerline{
\includegraphics[height=8cm]{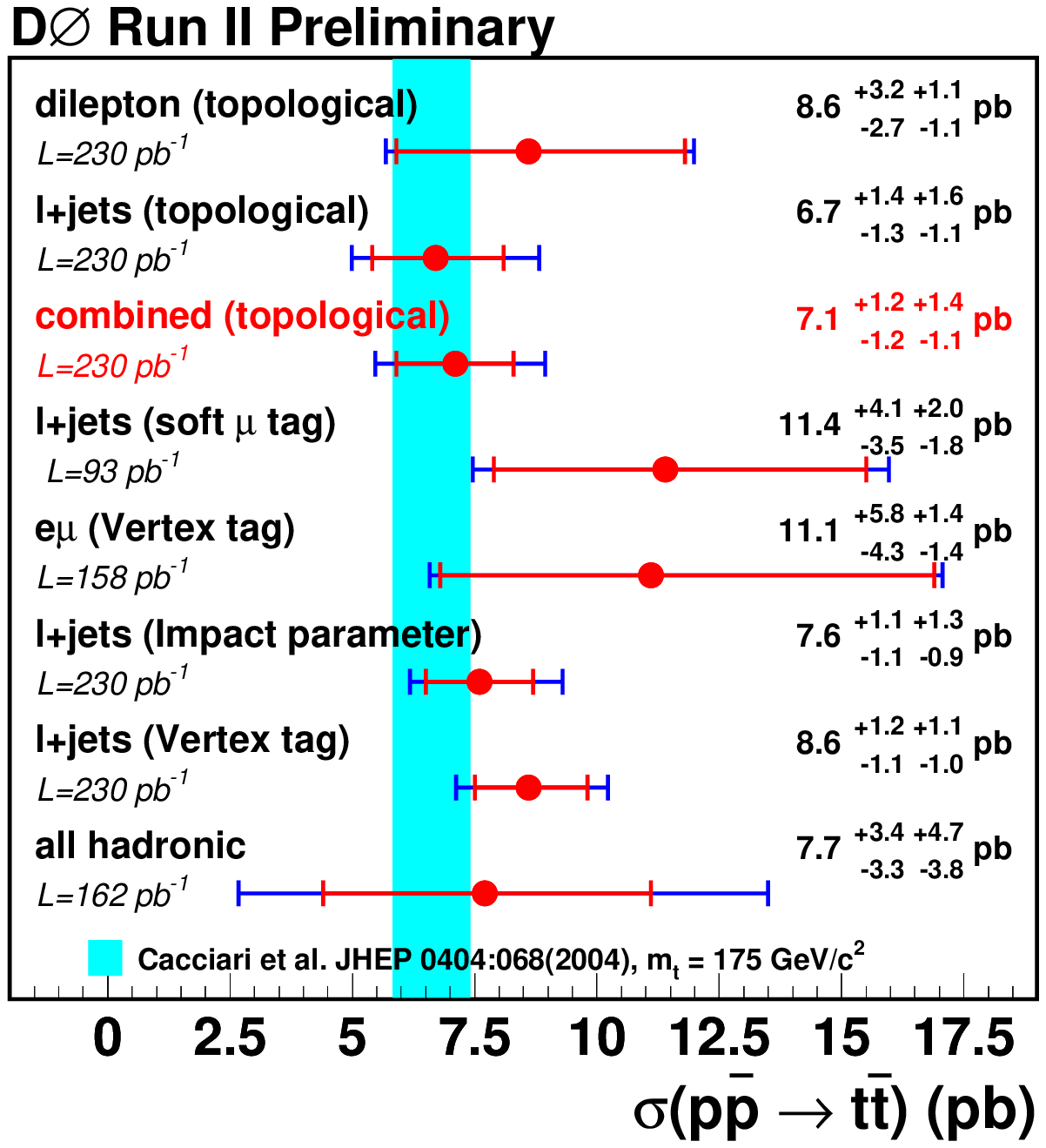}}
\caption[Top pair production at the Tevatron]{\label{sigmatt}
                Summary of cross sections for the production of top pairs
                as measured at the Tevatron in different final state topologies.
                    }
\end{minipage}
\end{figure}

W-pair production in e$^+$e$^-$ scattering is sensitive to ZWW and $\gamma$WW
triple gauge couplings. Another diagram contributing to W-pair production is
neutrino exchange. Because of the structure of the SM,
the couplings are in such a balance, that large cancellations occur in the cross
section for e$^+$e$^-$$\rightarrow$W$^+$W$^-$, which prevents run-away of this
cross section leading to violation of unitarity.
This is nicely demonstrated in Fig.~\ref{sigmaWW}, where the measurement
of all LEP experiments combined is compared to the SM, showing excellent
agreement.
In a slightly more sophisticated analysis the W-pair cross section and the
angular distribution of the W's can be used to derive the anomalous
gauge couplings $\kappa_{\gamma}$, $\lambda_{\gamma}$ and
$g_1^{\mathrm{Z}}$, which in the SM take the values 1, 0 and 1,
respectively. Results of fits for these anomalous couplings are shown
in Fig.~\ref{LEPTGC}. 
They are clearly in good agreement with the SM expectations.
Triple gauge boson couplings also play an important role in the
production of multiple gauge bosons in the same event at the
Tevatron. Figure~\ref{TeVmultiboson} shows the measurements
of the cross sections for single and multi boson production for
various combinations of bosons at the Tevatron.
The SM predictions are superimposed and show good agreement
with the measurements.

\section{Top quark properties}
\vspace*{-3mm}
At the moment the Tevatron is the only accelerator that produces top quarks.
Top quarks can be produced in pairs through the strong interaction
and singly in weak processes.
Top decays dominantly to a b-quark and a W boson,
where the W decays again into hadrons or leptons. 
For $\mathrm{t}\overline{\mathrm{t}}$ events this leads to
six experimental topologies: {\em di-lepton events} with a final state of
$2\ell+2\mathrm{b}$+lots of missing transverse energy;
{\em lepton+jets events} with a final state of $1\ell+4$~jets of which
2~jets are from b quarks and some missing transverse energy and;
{\em all-jets events} which consist of six jets in the final state of
which 2 are from b quarks.

The $\mathrm{t}\overline{\mathrm{t}}$
production cross section at the Tevatron is shown
in Fig.~\ref{sigmatt}.
More details are given in the QCD contribution to these
proceedings~\cite{Greenshaw}.
Single top production,
important to measure the $V_{\mathrm{tb}}$ element of the
Cabbibo-Kobayashi-Maskawa matrix (see~\cite{Branco,Shune}),
has not yet been observed by CDF and
D\O\ at the Tevatron. Upper limits are given of
13.6~pb (CDF~\cite{CDFsingletop}) and 6.4~pb (D\O~\cite{D0singletop})
for $s$-channel production (W decay) and
10.1~pb (CDF~\cite{CDFsingletop}) and 5.0~pb (D\O~\cite{D0singletop})
for $t$-channel production (W exchange) all at 95\% CL.
This is in agreement with the SM model expectations of
0.88$\pm$0.07~pb and 1.98$\pm$0.21~pb for $s$- and
$t$-channel respectively.
The Tevatron limits are typically based on data
corresponding to 200--250~pb$^{-1}$ of luminosity and a measurement
may be expected in the next year or so.

\begin{figure}[tbp]
\begin{minipage}[t]{7.25cm}
\centerline{\includegraphics[height=7.5cm]{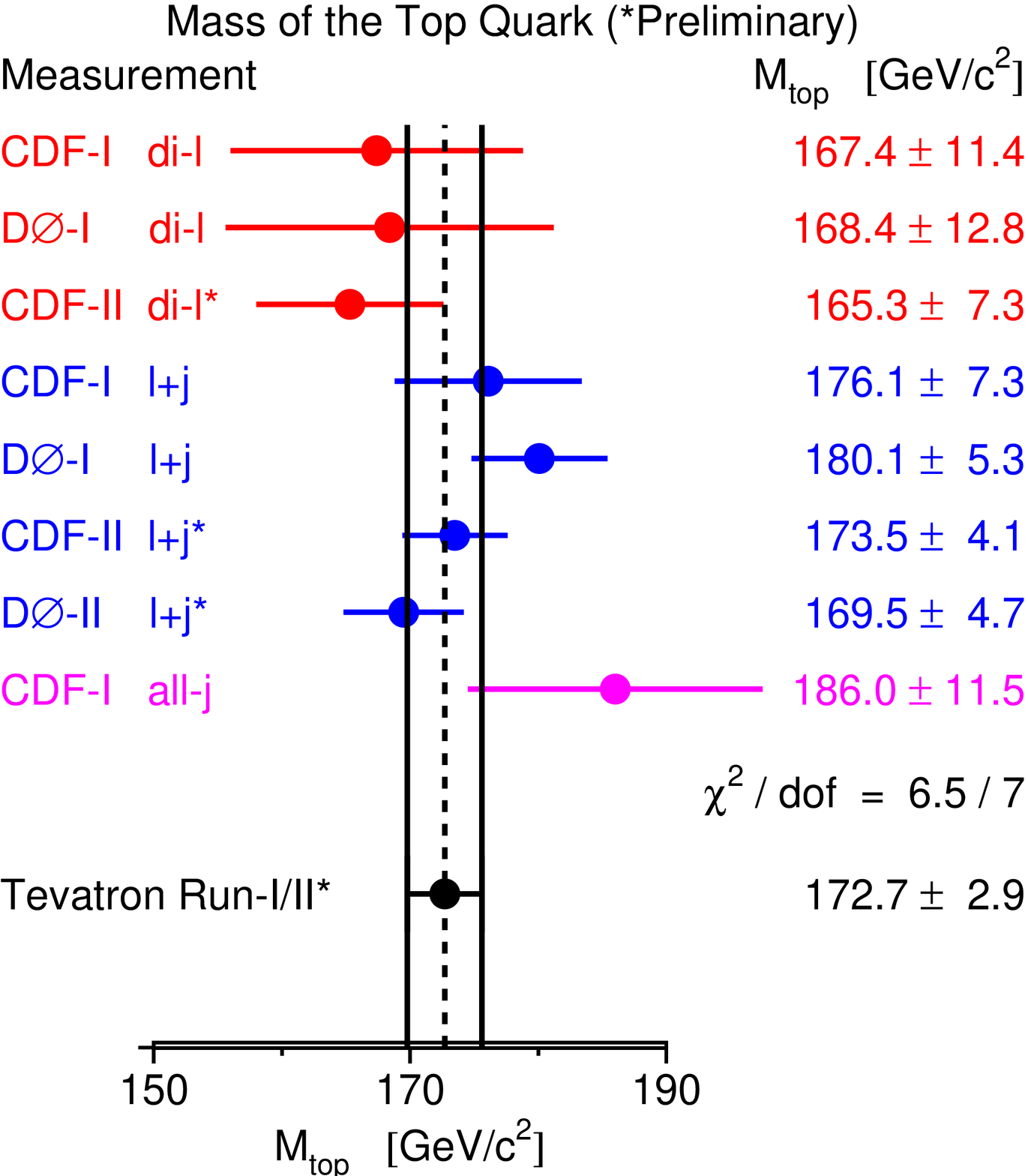}}
\caption[Top mass]{\label{mtop}
                Measurements and average of the top quark mass.
                    }
\end{minipage}
\hfill
\begin{minipage}[t]{7.25cm}
\centerline{
\includegraphics[height=8cm]{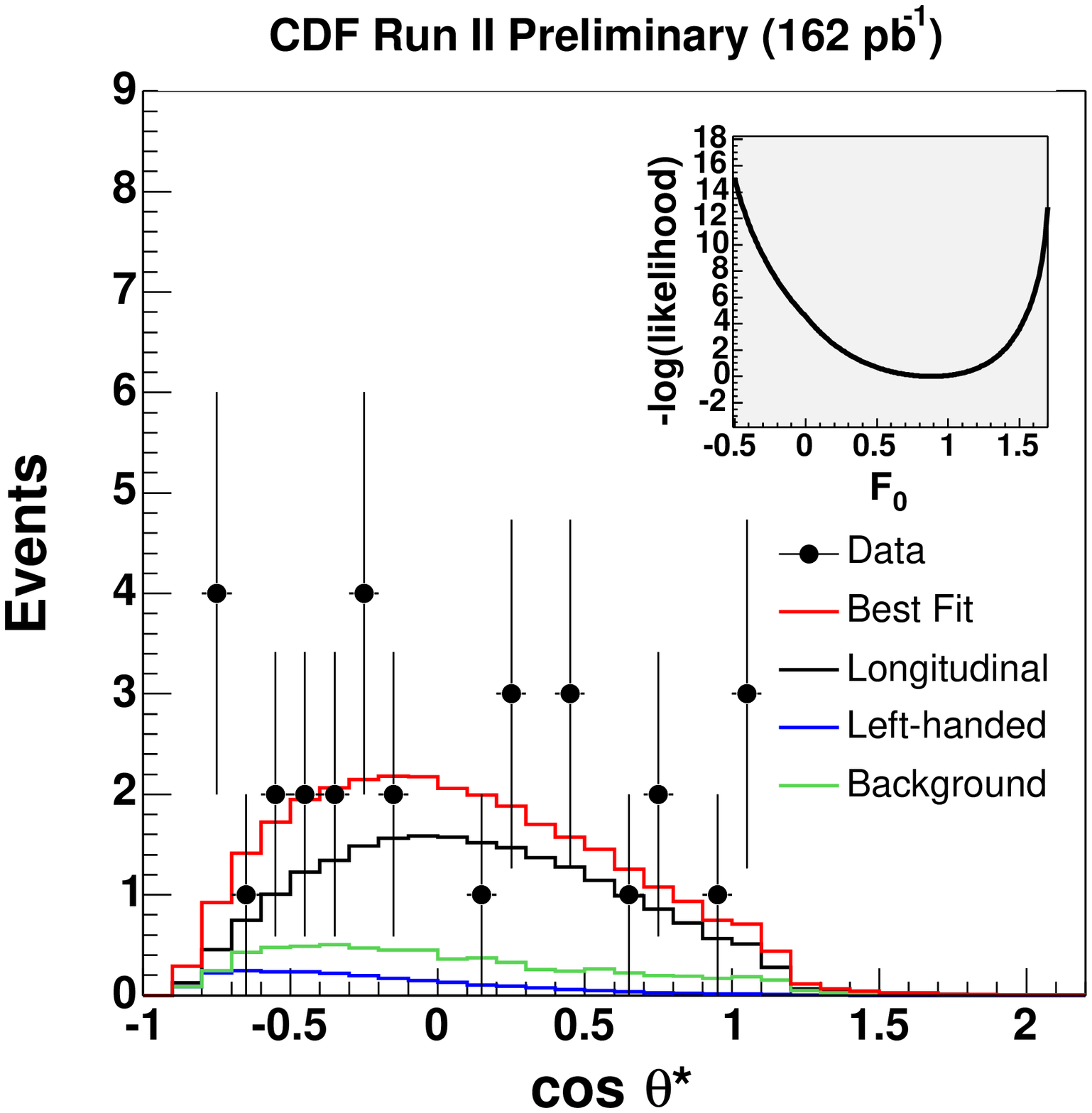}}
\caption[CDF cos(thetastar) for top events]{\label{costhetastar}
                CDF measurement of the angle between the W and
                b quark in a top quark decay
                $\cos\theta^*\approx 2m_{\ell\mathrm{b}}^2/\!\!\!\left(
                m_{\mathrm{top}}^2-m_{\mathrm{W}}^2\right)-1$.
                    }
\end{minipage}
\end{figure}

An important property to measure is the top quark mass.
The most important channel to do this is lepton+jets.
The di-lepton channel has very small statistics, while
the all jets channel suffers severe background.
Event selection for the lepton+jets channel is done by
identifying an electron or muon and four jets,
followed by a topological selection and b-tagging.
The signal to background ratio that can be observed is good, especially
after b-tagging.
The top mass is extracted by comparing observables to matrix elements
and MC predictions, either via templates or another a priori probability
density.
The signal probability can be determined from theory by
convoluting the cross section with a transfer function
that models how the observables that the cross section depends on
get smeared by fragmentation, detector resolution and analysis (such as
jet finding method):
CDF uses $m_{\mathrm{top}}$ directly as an observable,
while D\O\ uses more elementary observables, such as lepton and jet
energy and angles.
These methods allow to simultaneously fit an overall
Jet Energy Scale (JES) using the W-mass in the top-events
to constrain it.
This greatly reduces the uncertainty of the JES at the cost of a
larger statistical error.
The results of the analyses are listed in Fig.~\ref{mtop}.
Note the small systematic on the new results. The current preliminary
top mass is $m_{\mathrm{top}}=172.7\pm2.9$ GeV.
Both experiments expect to go under an error of 2 GeV eventually
in the Tevatron Run 2.
As pointed out by Gr\"unewald~\cite{Grunewaldparallel}
there might be a systematic trend in the top quark mass determinations
depending on the decay topology, which may become important as the
determination becomes more accurate.

Because the top quark decays before it hadronises, its helicity can be
measured through the angular distribution of its decay products.
The angle between the W and the b quark from the top decay in
the top rest frame can be approximated as
$\cos\theta^*\approx 2m_{\ell\mathrm{b}}^2/\!\!\!\left(
m_{\mathrm{top}}^2-m_{\mathrm{W}}^2\right)-1$.
In Fig.~\ref{costhetastar} the distribution of this angle $\cos\theta^*$
is plotted as measured by the CDF collaboration.
The $\cos\theta^*$ distribution is fitted to three helicity components:
$f_0$ with W direction along the top spin, W spin transverse to the top spin
and b quark spin along the top spin;
$f_-$ with W direction opposite the top spin, W spin along the top spin and
b quark spin opposite to the top spin; and
$f_+$ with W direction along the top spin, W spin along the top spin and
b quark spin opposite to the top spin.
The SM expectations for these quantities are
$f_0\approx 0.7$, $f_-\approx 0.3$ and
(due to the purely $V-A$ coupling) W $f_+=0$.
Also other angular information can be used in a similar way,
such as the transverse
moment of the lepton from the W decay and the invariant mass of the
lepton from the W decay and the b quark from the top decay.
From the fit to the $\cos\theta^*$ distribution CDF determined
$f_0=0.89_{-0.34}^{+0.30}\pm 0.17$~\cite{CDFf0-1}.
Another CDF measurement uses a combination of  di-lepton channel
and the lepton+jets channel of t$\overline{\mathrm{t}}$ events to obtain
$f_0=0.27_{-0.21}^{+0.35}$~\cite{CDFf0-2}.
lepton+jets channel of t$\overline{\mathrm{t}}$ events.
and $f_+ < 0.18$ at 95\% CL~\cite{CDFfplus},
whereas D\O\ determined that $f_+=0.04\pm 0.11\pm 0.06$
in the combination of the di-lepton and lepton plus jets
channels~\cite{D0topwhel}.
These results are all fully compatible to the SM prediction.

Using the fact that in $\mathrm{t}\overline{\mathrm{t}}$ events
there is a b quarks from both tops in the final state
and by counting the number of zero, one or two b-tagged jets,
the ratio of top quarks decaying into a b quark over any top decay,
$R=\mathrm{B}(\mathrm{t}\rightarrow\mathrm{Wb})/
\mathrm{B}(\mathrm{t}\rightarrow\mathrm{Wq})\approx |V_{\mathrm{tb}}|^2$,
can be measured and thus $|V_{\mathrm{tb}}|$.
The results
give rise to the limits
$|V_{\mathrm{tb}}|>0.78$ by CDF~\cite{CDFVtb} and
$|V_{\mathrm{tb}}|>0.80$ by D\O~\cite{D0Vtb} both at the 95\% CL,
well compatible with $|V_{\mathrm{tb}}|\approx 1$.

\section{The EW fit and prediction of the SM Higgs mass}
\vspace*{-3mm}
\begin{table}[bp]
\hfill
\begin{tabular}{|r@{~=~}r@{\hspace*{1mm}}l@{\hspace*{-1mm}}l|}
\hline
$\Delta\alpha_{\mathrm{had}}$&0.02767&$\pm$ 0.00034 & \\
$\alpha_{\mathrm{s}}$               &0.1186   &$\pm$ 0.0026   & \\
$m_{\mathrm{Z}}$                       &91.1874&$\pm$ 0.0021   & GeV \\
$m_{\mathrm{top}}$                   &173.3     &$\pm$ 2.7          & GeV \\
$m_{\mathrm{H}}$                      &91          &{\Large $_{-32}^{+45}$} & GeV \\[1mm]
\hline
\end{tabular}
\hfill
\begin{tabular}{r|cccc}
Correlation & & & & \\
coefficients
& $\Delta\alpha_{\mathrm{had}}$
& $\alpha_{\mathrm{s}}$
& $m_{\mathrm{Z}}$
& $m_{\mathrm{top}}$ \\
\hline
$\alpha_{\mathrm{s}}$                                   & 0.01  &            &          &        \\
$m_{\mathrm{Z}}$                                           & -0.01 & -0.02 &          &        \\
$m_{\mathrm{top}}$                                        & -0.02 & 0.05 & -0.03 &        \\
$\log(m_{\mathrm{H}}/(1~\mathrm{GeV})$ & -0.51 & 0.11 & 0.07 & 0.52 \\
\end{tabular}
\hfill
\caption[EW fit results]{\label{ewfitresults}
                Results of the electroweak fit as performed by the LEP
                EW Working Group~\cite{Grunewaldparallel}.
                    }
\end{table}
Fitting all relevant LEP, SLD and Tevatron
electroweak measurements simultaneously
as is done by the LEP ElectroWeak Working Group yields,
when using also the $\Delta\alpha_{had}$ from
Burkhardt and Pietrzyk, the outputs shown in Fig.~\ref{LEPEWtable}.
and Table~\ref{ewfitresults}
The only SM parameter without direct experimental determination is
the SM Higgs boson mass, $m_{\mathrm{H}}$.
The fit gives an indirect determination of
log$(m_{\mathrm{H}})=1.96\pm0.18$.
The fit has a $\chi^2/$d.o.f.$=17.8/13$ corresponding to a fit
probability of 16.5\%.
Looking at the correlation matrix between the fit parameters
also given in Table~\ref{ewfitresults},
it becomes clear why above relatively much attention was given to
$\Delta\alpha_{had}$ and $m_t$
in view of the importance to determine $m_{\mathrm{H}}$.
\begin{figure}[btp]
\begin{minipage}[t]{7.25cm}
\centerline{\includegraphics[height=7.5cm]{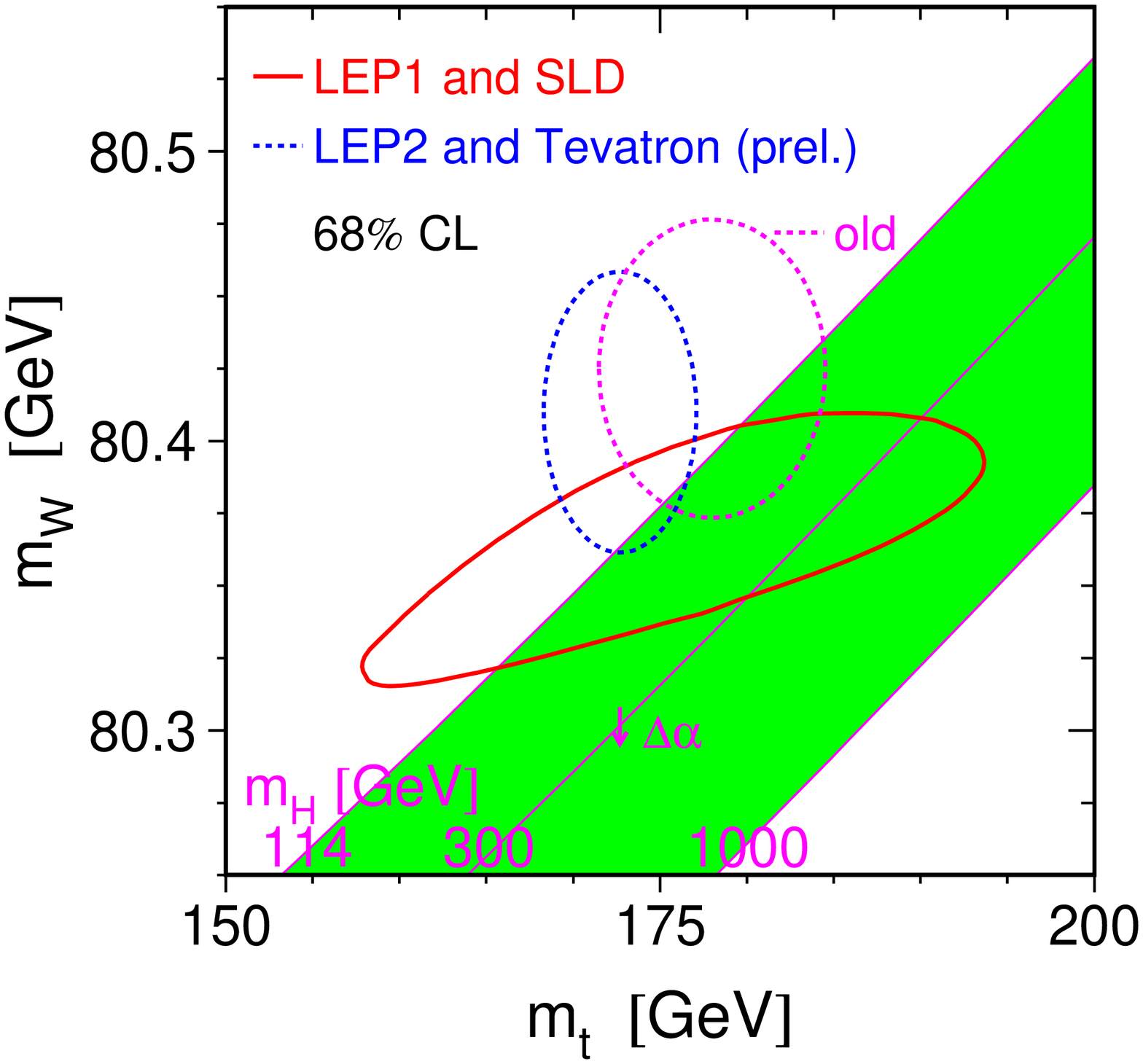}}
\caption[Top mass versus W mass]{\label{mtmwhiggs}
                Probability contours (ellipses) at 68\% CL of the EW fit in the
                $m_{\mathrm{top}}$--$m_{\mathrm{W}}$ plane.
                The ellipse indicated as ``old'' is the one that was presented at the
                Lepton-Photon conference in summer 2005.
                The grey area is the SM prediction for various Higgs boson masses,
                that label de diagonals.
                    }
\end{minipage}
\hfill
\begin{minipage}[t]{7.25cm}
\centerline{
\includegraphics[height=8cm]{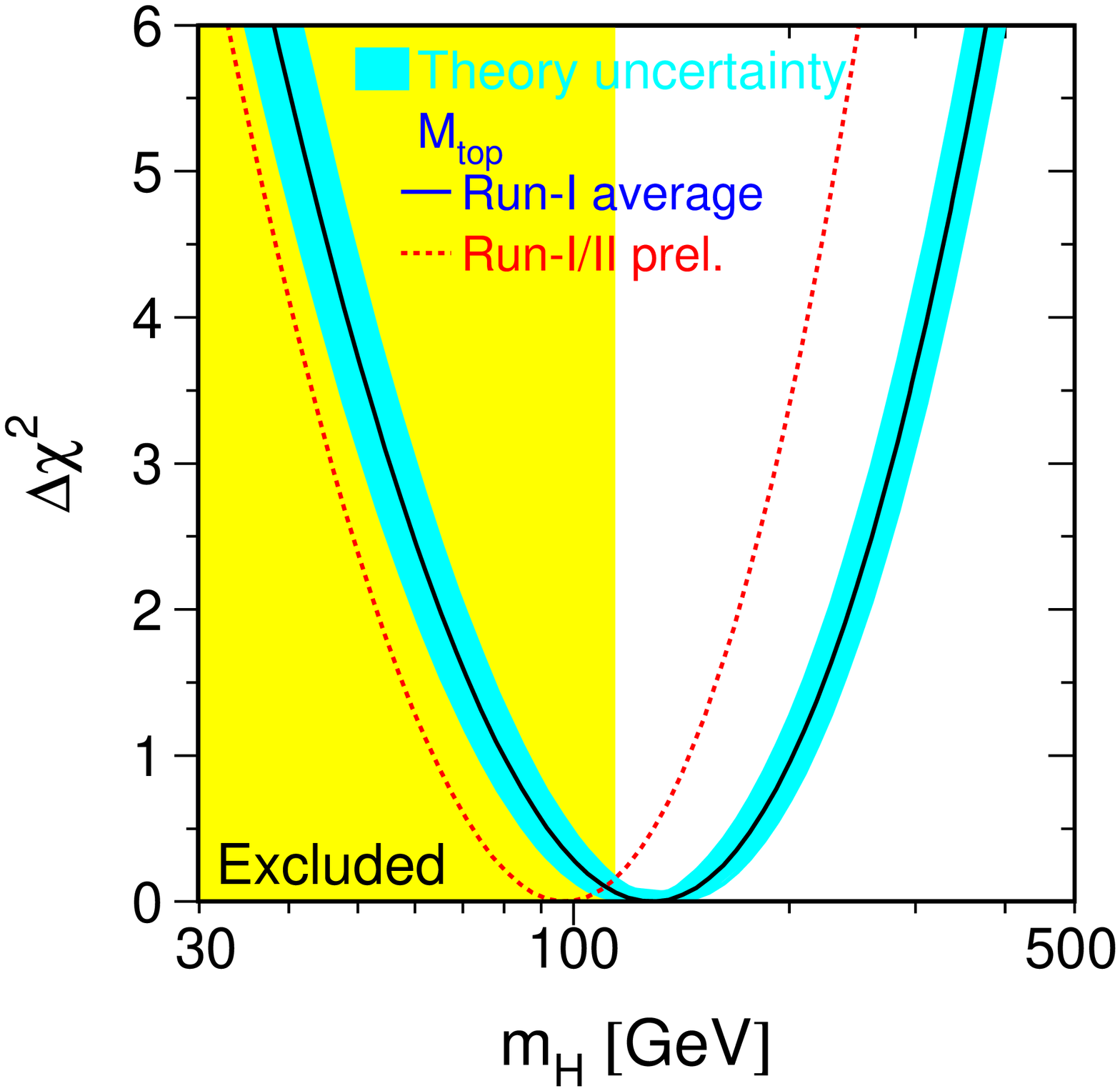}}
\caption[Blueband plot]{\label{blueband}
               Difference in $\chi^2$ with respect to the best fit, as a function
               of the Higgs boson mass.
               The dotted curve is the current preliminary fit for which the
               theoretical error has not yet been established.
               The band is an older fit and can be used as an indicator for the
               theoretical uncertainty.
               In light grey the Higgs boson mass range that has been excluded
               by direct searches at LEP.
                    }
\end{minipage}
\end{figure}
Comparing $m_{\mathrm{}W}$ and $m_{\mathrm{top}}$ from direct and indirect measurements to the SM expectation in Fig.~\ref{mtmwhiggs},
a consistent picture arises with a preference for a low Higgs mass.
Compared to last year, notably the $m_t$ measurement improved considerably,
but the qualitative conclusion has hardly changed.
In Fig.~\ref{blueband} the situation for the prediction of the Higgs boson mass
is summarised. When the probability from the $\chi^2$ values is normalised to
only include the Higgs mass range above the direct search limit (see next section)
as possibilities, an upper limit for the SM Higgs boson mass can be derived of
$m_{\mathrm{H}}<219$~GeV at 95\% CL.

\section{Direct search for the SM Higgs}
\vspace*{-3mm}
At LEP-2 the Higgs boson has been primarily searched for in the production
channel in which it is radiated off a virtual Z boson.
The fact that no clear signal for Higgs boson production has been observed
at LEP-2 has lead to a lower limit of the possible mass of the Higgs boson
of $m_{\mathrm{H}}>114.4$~GeV at 95\% CL by combining the searches in
all possible final state channels by all LEP experiments~\cite{LEP2Higgs}.
Although this limit is still preliminary, it is quite stable for a while now.

\begin{figure}[tp]
\vspace*{-5.5mm}
\begin{minipage}[b]{5,5cm}
\centerline{\includegraphics[height=4cm]{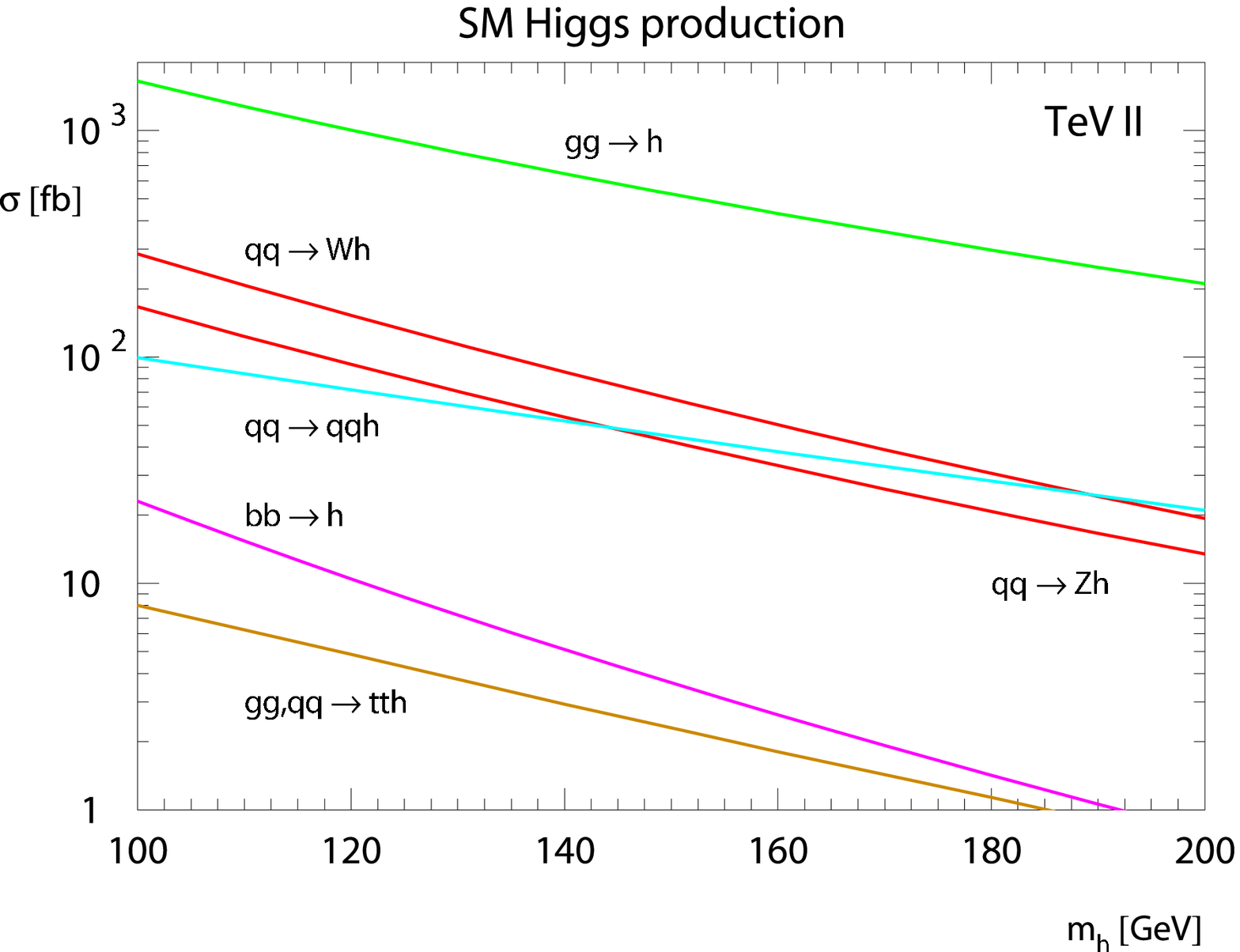}}
\caption[Higgs cross section at Tevatron vs mass]{\label{tevhiggsxs}
                Cross section for SM Higgs production at the Tevatron at
                $\sqrt(s)=1.96$~TeV as a function of the Higgs mass
                in the various production channels.
                    }
                    \vspace*{2mm}
\end{minipage}
\hspace{3mm}
\begin{minipage}[b]{3.5cm}
\centerline{
\includegraphics[height=4cm]{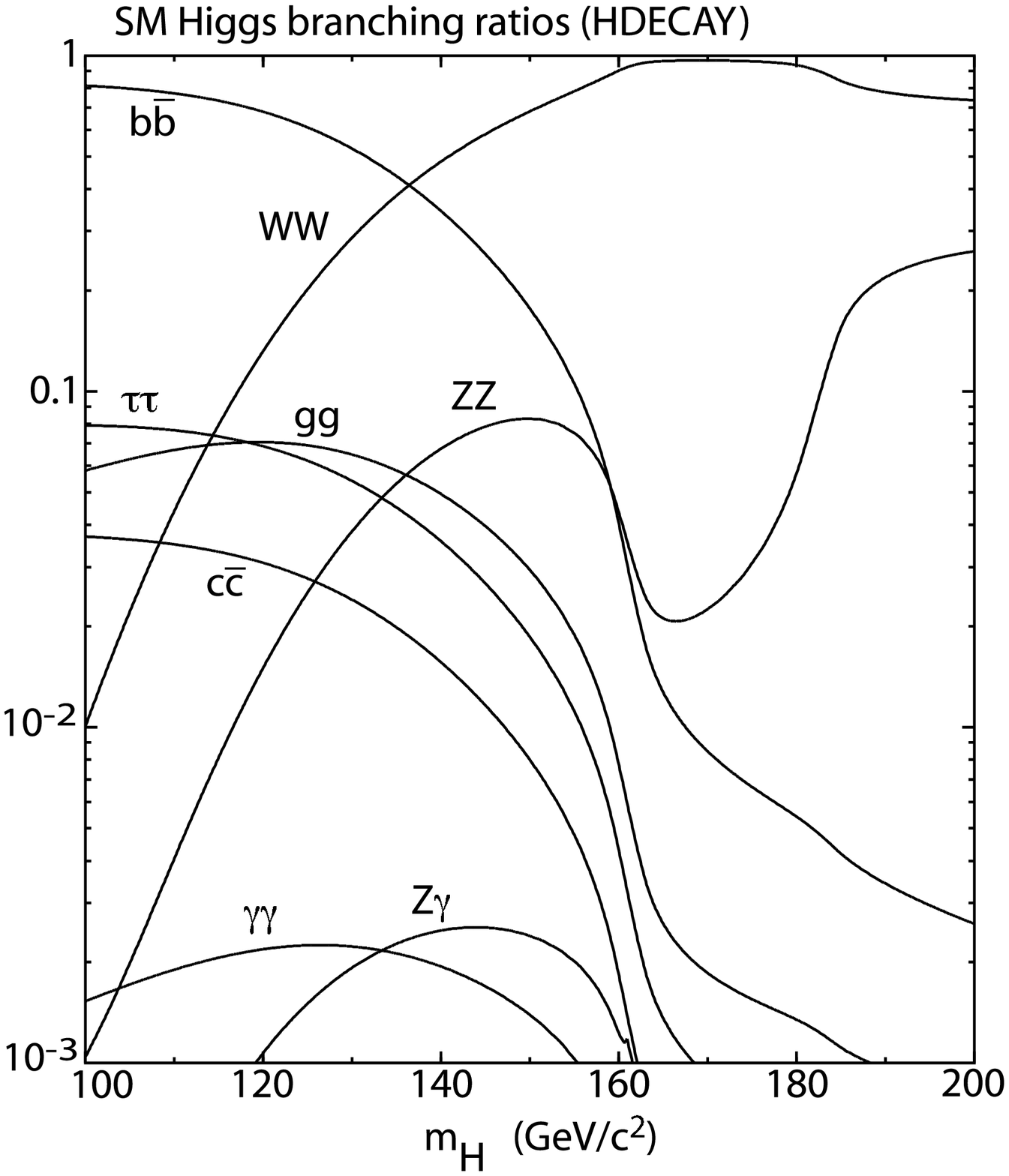}}
\caption[Higgs branching ratio vs mass]{\label{tevhiggsdecay}
               Branching ratio of the SM Higgs boson as a function of their mass
               in the various decay channels.
                    }
                    \vspace{2mm}
\end{minipage}
\hfill
\begin{minipage}[b]{5cm}
H decay is up to $m_{\mathrm{H}}=135$~GeV
dominantly into
b$\overline{\mathrm{b}}$
and above that into WW$^{(*)}$ (where one of the W's is often off-shell).
Most of the region of interest is in the overlap region where both
decays have a non-negligible branching fraction.
H$\rightarrow$b$\overline{\mathrm{b}}$
has large backgrounds and  is easiest searched for when the Higgs is
produced in association with a W or Z that decays leptonically,
giving a charged lepton and/or missing $E_{\mathrm{T}}$ trigger.
Lifetime or
\end{minipage}
\vspace*{-6.5mm}
\end{figure}

The current best place to look for a Higgs boson in the mass range from
114 to 219~GeV is at the Tevatron.
%
In Fig.~\ref{tevhiggsxs} and~\ref{tevhiggsdecay} the cross section for SM
Higgs production in the main channels and the branching ratios for the
decay of the SM Higgs boson are plotted.

\noindent
soft lepton b quark tagging is used to identify the jets from H decay.
The background mostly consists of W or Z plus jets and is getting
progressively better understood from the data and Monte Carlo.
The expected signal to background ratio is not very large.
H$\rightarrow$WW$^{(*)}$ can be selected in the channel where
one or both W's decay leptonically to electron or muon.
A promising channel is WH production followed by H$\rightarrow$WW$^{(*)}$,
where of the 3W's produced the like-sign pair is selected through leptonic decay,
allowing a very clean selection.

\begin{figure}[bp]
\vspace*{-3mm}
\begin{minipage}[t]{7.4cm}
\centerline{\includegraphics[height=5cm]{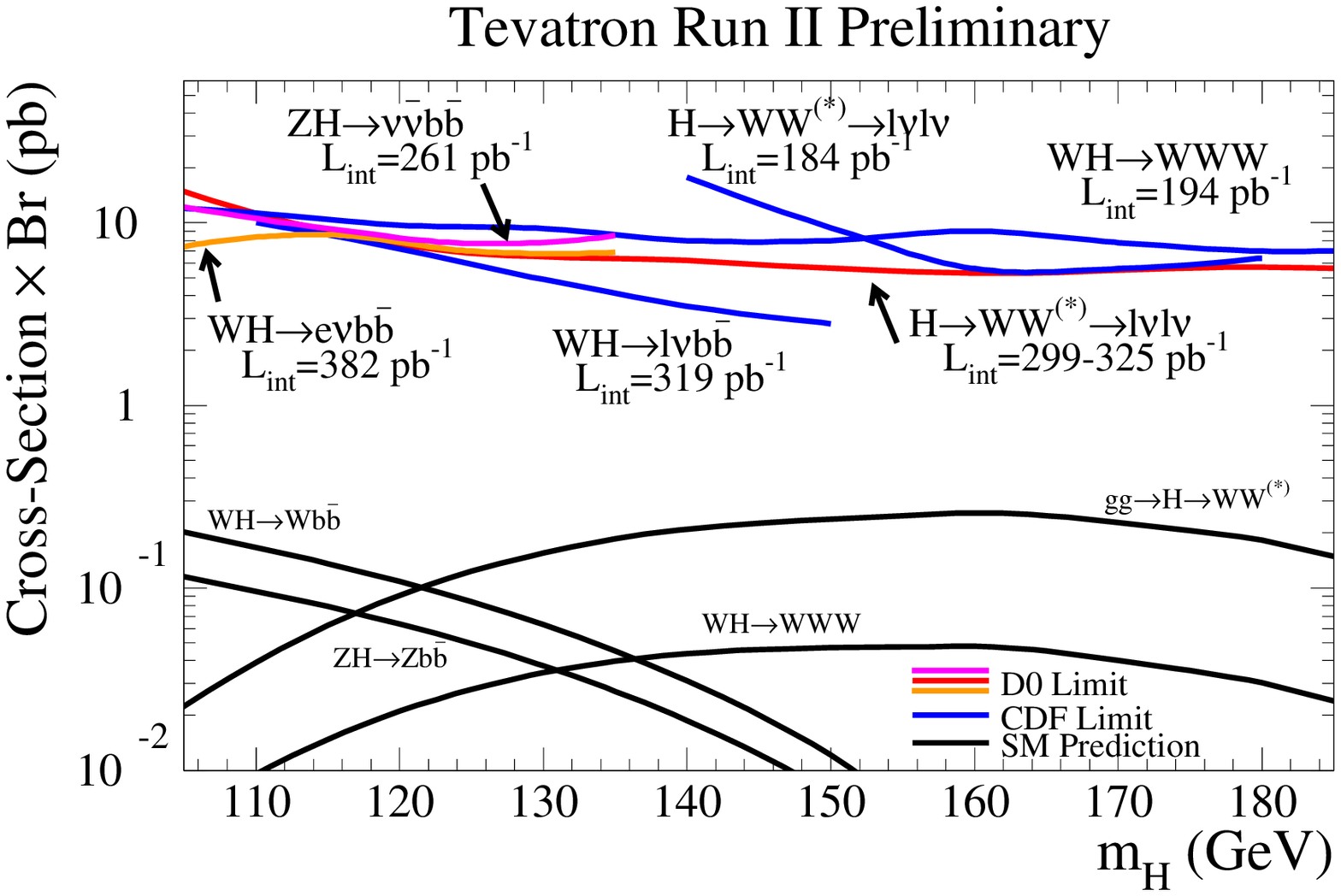}}
\caption[Higgs cross section times branching ratio limits at the Tevatron vs mass]
                {\label{TeVHiggs}
                Cross section times branching ratio upper limits at 95\% CL
                for Higgs production at the Tevatron at
                $\sqrt(s)=1.96$~TeV as a function of the Higgs mass
                in the various search channels as indicated for the curves in the
                upper part of the figure~\cite{TeVHiggs}.
                In the lower part of the figure the SM cross section times
                branching ratio are given for the relevant production and
                decay topologies.
                }
\end{minipage}
\hfill
\begin{minipage}[t]{=7.4cm}
\centerline{
\includegraphics[height=5cm]{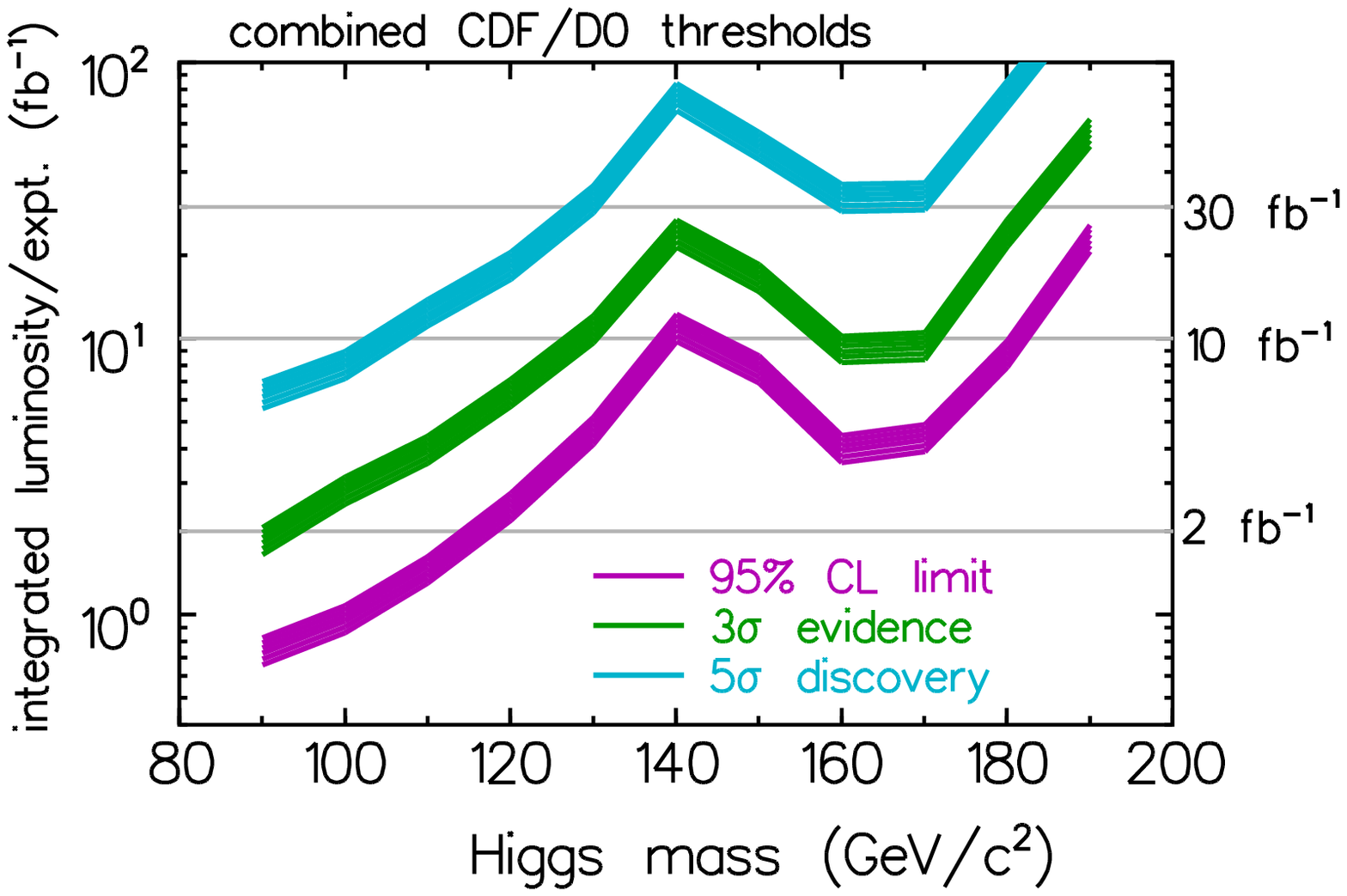}}
\caption[TeV2000 Higgs reach]{\label{TeV2000}
               Summary of exclusion or discovery range as a function of the
               Higgs mass for the Tevatron Run 2, as reported by the Tevatron
               Higgs Working Group in 2000~\cite{TeV2000}. 
               The lower band indicates the average luminosity for which a
               95\% CL limit on the existence of the Higgs can be obtained
               as a function of the Higgs mass.
               The middle and upper curve show the average luminosity at
               which a three or five standard deviation discovery can be
               made as a function of the Higgs mass. 
                    }
\end{minipage}
\end{figure}
Combining all current, mostly preliminary results, the picture
arises as shown in Fig.~\ref{TeVHiggs}.
It is clear from this figure that the present sensitivity is at least an order
of magnitude away from being
able to exclude the existence of the Higgs
boson at mass ranges above the values excluded by the
direct searches at LEP.
In 1999 and 2000 a study was made of the Tevatron Run 2 sensitivity
to exclude or discover the Higgs. The predicted result is shown in
Fig.~\ref{TeV2000}.
Comparing the current cross section limit results in Fig.~\ref{TeVHiggs}
to the Working Group expectations, the
H$\rightarrow$b$\overline{\mathrm{b}}$
part is currently nearly a factor of 10 under the expected sensitivity.
However  improvements by a factor 4 in b-tagging and another factor of
$\sqrt{2}$ by using both
the electron and muon 
decay channels of the W can
be attained. 
Another factor of up to $\sqrt{2}$ can be reached by considering
a larger range in geometric acceptance than only the central region
presently considered.
These are the first generation of results for these searches and additional
gain in sensitivity by refining the search strategy can be expected.
All in all the projected sensitivity from the Working Group back in 2000
seems attainable.

In the H$\rightarrow$WW$^{(*)}$ regime we are still a factor of 2
under the expected sensitivity.
But also here improvements are worked on, although a single
improvement that gives a large part of this factor cannot easily be
identified and the
gain is likely to come from a number of small improvements.

There was an update of the sensitivity prediction in 2003, showing a
larger sensitivity~\cite{TeV2003}.
Attaining this sensitivity seems only possible in an optimistic scenario.

\section{Prospects for Higgs discovery in the next few years}
\begin{figure}[tp]
\begin{minipage}[t]{7.8cm}
\centerline{\includegraphics[height=4.5cm]{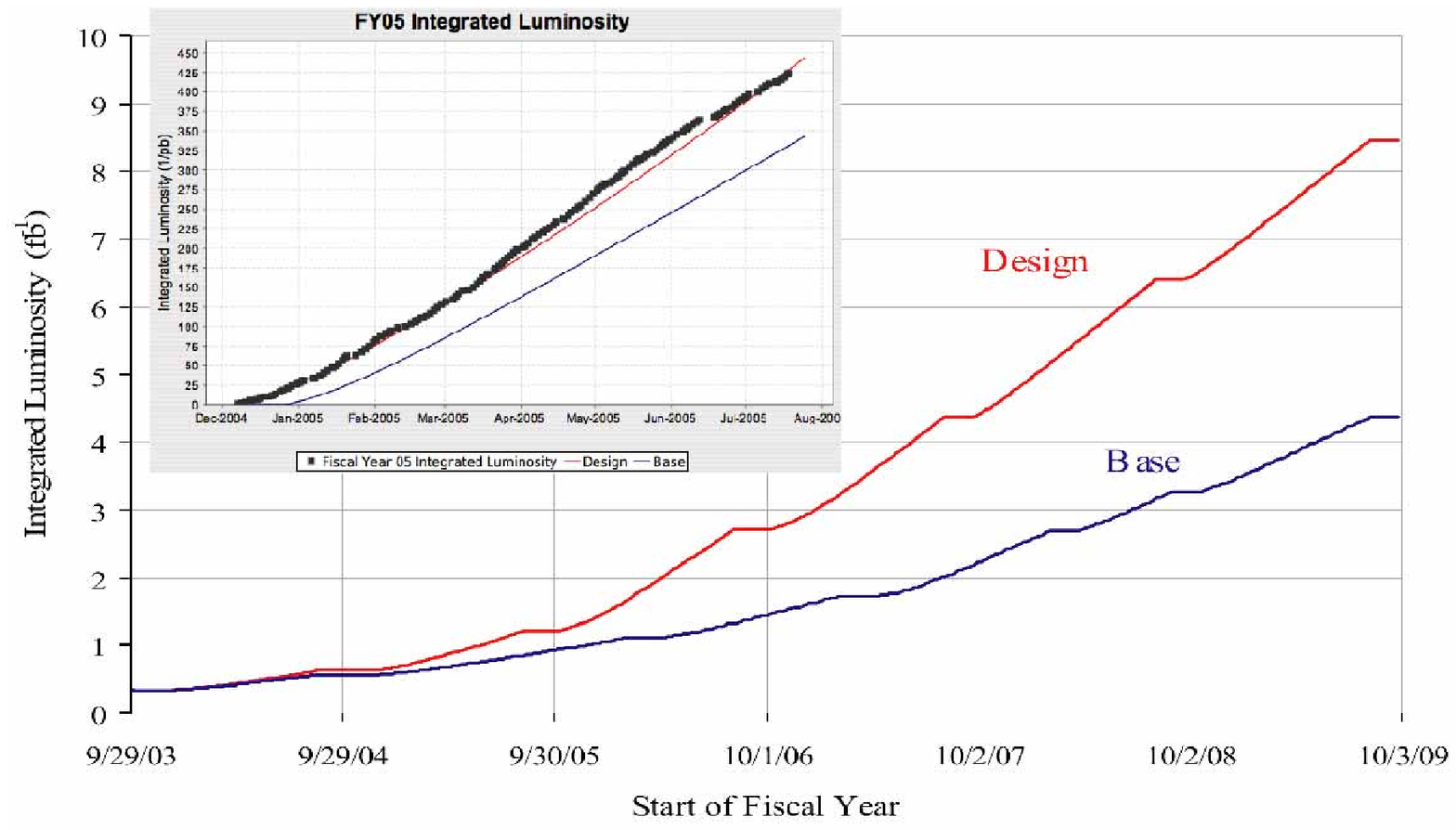}}
\caption[Luminosity prospects for the Tevatron]
                {\label{TeVlumi}
                Prospects for the baseline and design integrated luminosity
                per experiment for the Tevatron~\cite{TeVlumi}.
                The insert shows in more detail the integrated luminosity
                prospects and achievements up to the date of the conference.
                }
\end{minipage}
\hfill
\begin{minipage}[t]{=7.0cm}
\centerline{
\includegraphics[height=4.5cm]{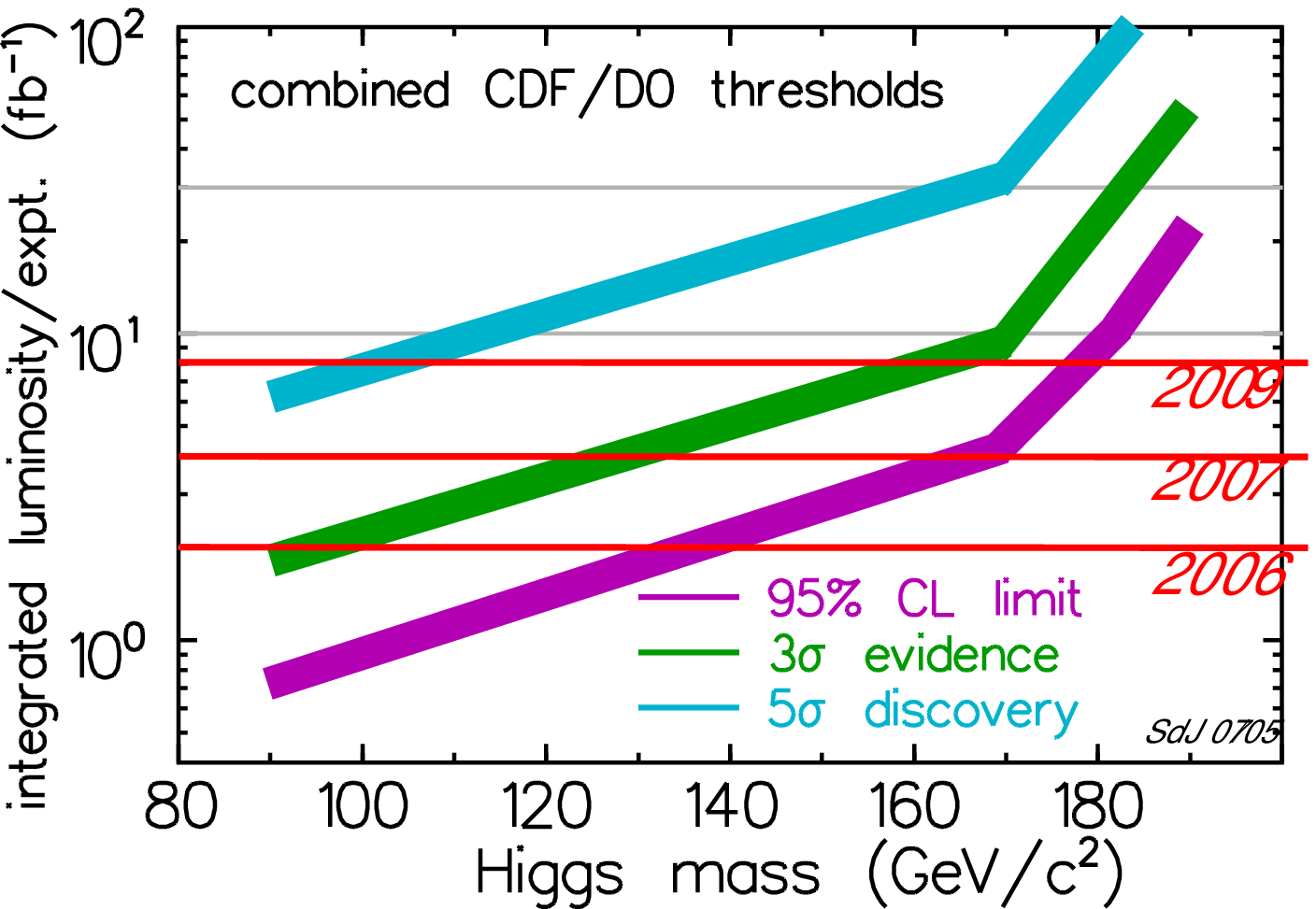}}
\caption[TeV2000 Higgs reach]{\label{TeV2000new}
               Version revised by the author of the exclusion or discovery
               range as a function of the Higgs mass for the Tevatron Run 2.
               The coding of the bands is as in Fig.~\ref{TeV2000}.
               On the right hand side the year is indicated when a certain
               integrated luminosity is taken as predicted in Fig.~\ref{TeVlumi}.
               }
\end{minipage}
\vspace*{-4mm}
\end{figure}

\vspace*{-3mm}
A striking feature in Fig.~\ref{TeV2000} is the middle of this plot which
is shared by all three curves.
This bump is in the overlap regime between the Higgs dominantly
decaying to b$\overline{b}$ or WW$^{(*)}$.
However, in the figure describing the actual measurement,
Fig.~\ref{TeVHiggs}, the cross section limit has a smooth behaviour,
even after the improvements mentioned before. Also the SM cross
section times branching ratio limits shown in this same
figure are approximately flat over the whole relevant Higgs
mass range.
It is therefore reasonable to assume that the bump in the
prediction of Fig.~\ref{TeV2000} will in reality not occur.
Assuming that in the pure
H$\rightarrow$b$\overline{\mathrm{b}}$ and
H$\rightarrow$WW$^{(*)}$
the sensitivity indicated in Fig.~\ref{TeV2000} will be attained
and the results are carefully combined in the intermediate overlap region,
the prediction for the sensitivity becomes that of Fig.~\ref{TeV2000new}.
In this figure the luminosity levels are indicated that should
be attained according to the Tevatron design luminosity projections at the
end of 2006 (2~pb$^-1$), the end of 2007 (5~pb$^-1$) and by the summer
of 2009 (8~pb$^-1$), according to the predictions in Fig.~\ref{TeVlumi}.
It should be noted that thus far the Tevatron has delivered a luminosity
equal to or exceeding the design values and
several upgrades to the Tevatron are planned
to maintain this trend.

\begin{figure}[tp]
\vspace*{-3mm}
\includegraphics[height=7cm]{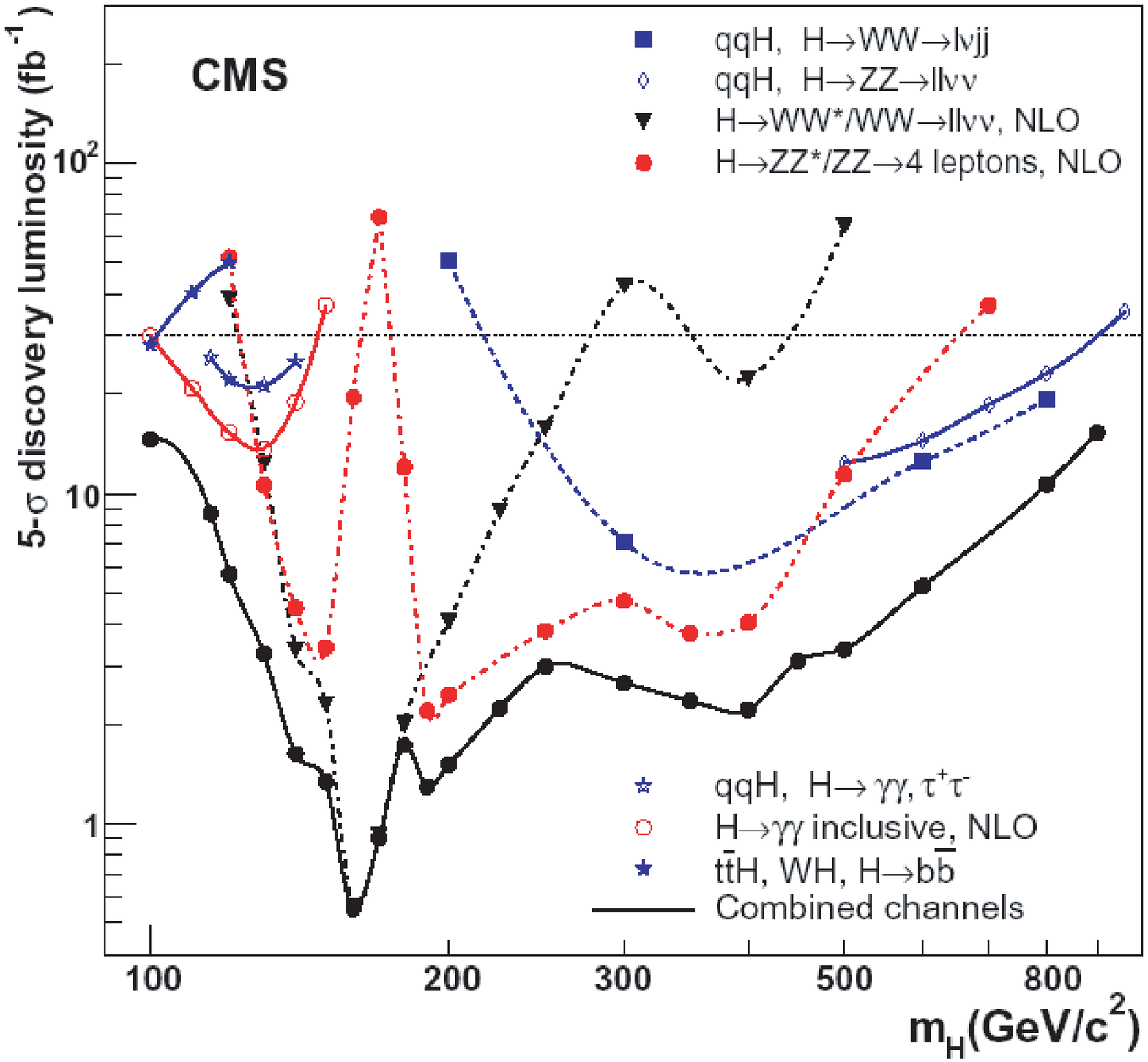}
\hfill
\includegraphics[height=7cm]{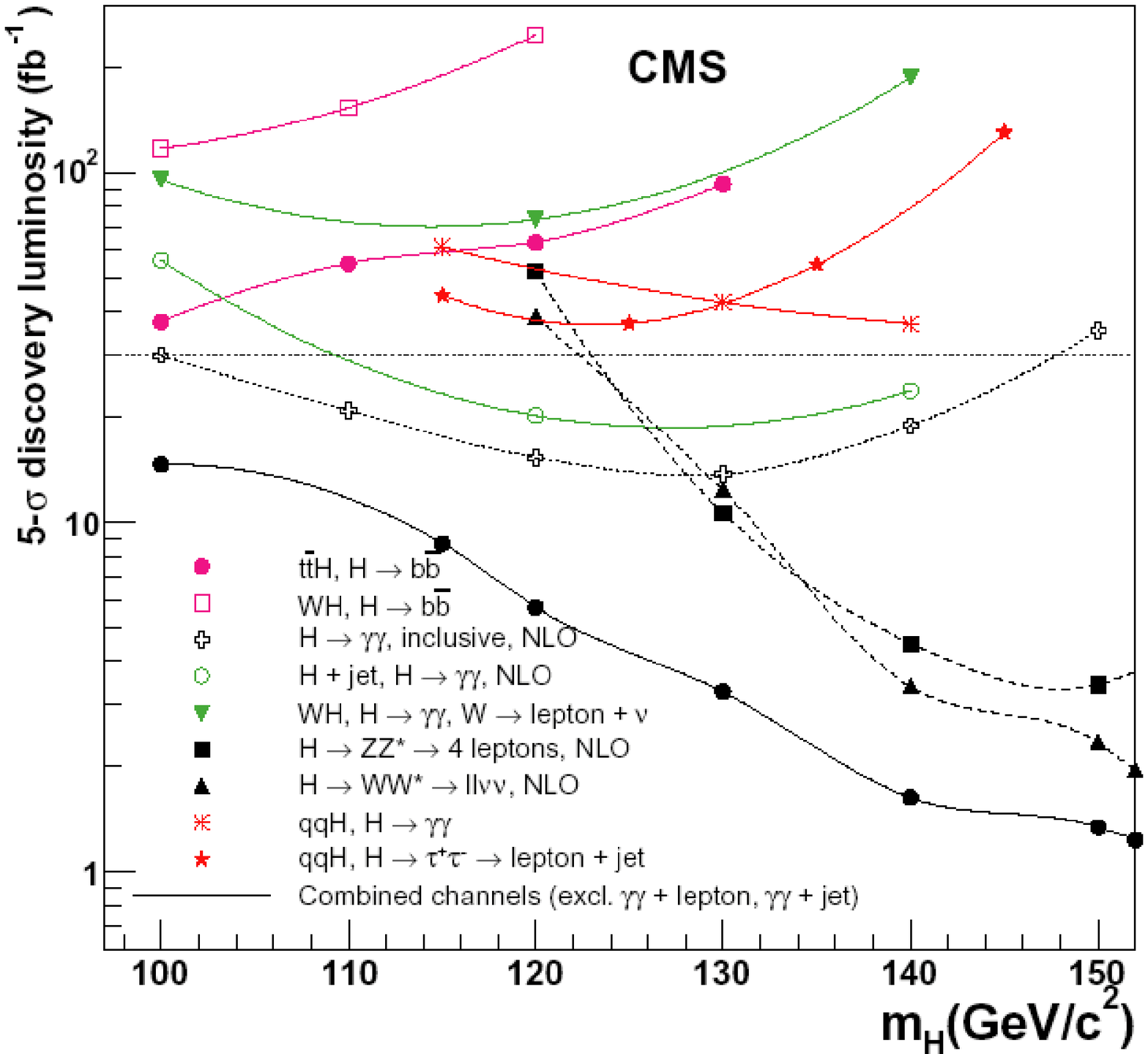}
\caption[CMS Higgs reach]{\label{CMSHiggs}
               Prediction for the luminosity needed to discover the SM Higgs
               boson with the CMS detector at the LHC as a function of the
               Higgs mass~\cite{dHondt}.
               }
\end{figure}
Clearly the next stop after the Tevatron to discover the Higgs boson and
measure its properties is the LHC.
In Fig.~\ref{CMSHiggs} the discovery potential for the SM Higgs boson
is given for the CMS experiment. The ATLAS experiment has very similar
sensitivity.
In this figure the approximate integrated luminosities
that correspond with one and two years of running
are also indicated.
If the LHC turns on as scheduled in 2007 and serious luminosity acquisition
is taken starting early 2008, discovering the SM Higgs will be a photo-finish
between the LHC and the Tevatron.
Taking the end of 2007 as a benchmark, the Tevatron may have excluded the SM
Higgs up to 160 GeV. This is excellent news for the LHC:
the SM Higgs is right where the LHC experiments are most sensitive,
where the Higgs branching ratio to W- and Z-pairs is large.
Alternatively, a hint of a Higgs with mass below 120 GeV may be present:
Again good news for the LHC: the SM will break down in the TeV range,
and the LHC will probably find signals of physics beyond the SM.

\section{Summary and conclusion}
\vspace*{-3mm}
The electroweak sector of the SM is in excellent overall agreement
with the available body of measurements.
The parts that are least in accordance with the SM predictions are the
lepton asymmetry from SLD measurements and the effective $\sin^2\theta_w$
measurement from NuTeV.
Improvement in the determination of the hadronic correction to
$\alpha_{\mathrm{QED}}$ is needed to resolve possible discrepancies
to describe the muon anomalous magnetic moment.
Improvements in $\alpha_{\mathrm{QED}}(M_{\mathrm{Z}})$,
$m_{\mathrm{W}}$ and $m_{\mathrm{top}}$ will lead to a
better estimate of the Higgs boson mass,
the only particle from the SM that has not yet been observed.
The precision of the measurements and SM predictions allow a mass range
$114<m_{\mathrm{H}}<219$~GeV at better than 95\% Confidence Level.
A Standard Model Higgs in this mass range will likely be observed
in the years 2007--2010, either at the Tevatron or at the LHC.

\newpage
\section*{Acknowledgements}
\vspace*{-3mm}
The results presented here are the work of many people, experimenters
and theorists alike and the author has merely freely quoted their results.
In particular I have had help from
Frederic Deliot, Simon Eidelman,
Richard Hawkings, Aurelio Juste,
Martin Gr\"{u}newald, Sven Heinemeyer,
Dick Kellogg, Kevin McFarland, Salvatore Mele,
Emmanuelle Perez, Bolek Pietrzyk,
Guenther Quast, Rik Yoshida,
Pippa Wells and Matthew Wing.
Of course, any errors and inaccuracies in this write-up remain the responsibility of the
author.

\end{document}